\documentclass[aps,pre,amsmath,amsfonts,floatfix,superscriptaddress,nofootinbib]{revtex4}

\usepackage{t1enc,latexsym,amsfonts, amssymb, amsmath,empheq,amsthm,bm}
\usepackage{graphicx}
\usepackage{stmaryrd}

\usepackage{mathrsfs}
\usepackage[usenames,dvipsnames]{xcolor}
\usepackage{color}
\usepackage{multirow}
\usepackage{bbm}
\usepackage{commath}

\usepackage{tikz}
\usetikzlibrary{shapes,arrows,positioning,calc}
\usepackage{hyperref}

\usepackage[french,english]{babel}
\usepackage[utf8x]{inputenc}
\usepackage[T1]{fontenc}
\usepackage{times}
\usepackage{url}

\definecolor{darkgreen}{rgb}{0,0.6,0}
\definecolor{darkblue}{rgb}{0,0,0.6}
\definecolor{darkred}{rgb}{0.6,0,0}
\definecolor{darkpurple}{rgb}{0.5,0,0.5}

\hypersetup{
bookmarksopen=true,
pdftitle="Mean-field dynamics of infinite-dimensional particle systems: global shear versus random local forcing",
pdfauthor="E Agoritsas", 
pdftoolbar=false, % toolbar hidden
pdfstartview={FitH},		% fits the width of the page to the window
pdfmenubar=true,			%menubar shown
pdfhighlight=/O,			%effect of clicking on a link
colorlinks=true,			%couleurs sur les liens hypertextes
urlcolor=darkblue,
citecolor=darkblue,		%couleur des liens pour les citations
linkcolor=darkpurple,	%darkred %couleur des liens hypertextes internes
}

\newcommand{\argc}[1]{\left[#1\right]}
\newcommand{\arga}[1]{\left\lbrace #1\right\rbrace }
\newcommand{\argp}[1]{\left(#1\right)}
\newcommand{\valabs}[1]{\vert #1\vert}
\newcommand{\moy}[1]{\left\langle  #1 \right\rangle }

\def\de{\mathrm d}
\def\redv{\bar v}
\def\rr{\mathbf{r}}

\def\xx{\mathbf{x}}
\def\uu{\mathbf{u}}
\def\vv{\mathbf{w}}
\def\cc{\mathbf{c}}
\def\de{\mathrm d}

\begin{document}

\title{
Mean-field dynamics of infinite-dimensional particle systems:
\\
global shear \textit{versus} random local forcing
}

%_____________________________________________________________

\author{Elisabeth Agoritsas}
\affiliation{Institute of Physics, Ecole Polytechnique F{\'e}d{\'e}rale de Lausanne (EPFL), CH-1015 Lausanne, Switzerland}

%_____________________________________________________________

\begin{abstract}

In infinite dimension, many-body systems of pairwise interacting particles provide exact analytical benchmarks for features of amorphous materials, such as the stress-strain curve of glasses under quasistatic shear.
Here, instead of a global shear, we consider an alternative driving protocol as recently introduced in Ref.~\cite{morse_roy_agoritsas_2020_Arxiv-2009.07706},
which consists of randomly assigning a constant local displacement on each particle,
with a finite spatial correlation length.
We show that, in the infinite-dimension limit, the mean-field dynamics under such a random forcing is strictly equivalent to that under global shear, upon a simple rescaling of the accumulated strain.
Moreover, the scaling factor is essentially given by the variance of the relative local displacements on interacting pairs of particles,
which encodes the presence of a finite spatial correlation.
In this framework, global shear is simply a special case of a much broader family of local forcing, that can be explored by tuning its spatial correlations.
We discuss specifically the implications on the quasistatic driving of glasses
--initially prepared at a replica-symmetric equilibrium--
and how the corresponding `stress-strain'-like curves and the elastic moduli can be rescaled onto their quasistatic-shear counterparts.
These results hint at a unifying framework for establishing rigourous analogies, at the mean-field level, between different driven disordered systems.

\end{abstract}

%_____________________________________________________________

\maketitle

%%% Date
%\begin{center}
%\today
%\end{center}

%_____________________________________________________________
%% Table of contents

%\begin{center}
%\rule{200pt}{0.5pt}
%\end{center}
%
%\tableofcontents
%
%\begin{center}
%\rule{200pt}{0.5pt}
%\end{center}

%_____________________________________________________________________________________________________
\section{Introduction}
\label{section-introduction}

Theoretical descriptions of driven amorphous materials remain challenging,
despite of decades of extensive analytical and computational studies
\cite{arceri_landes_2020_Arxiv-2006.09725,berthier_biroli_2011_RevModPhys83_587,nicolas_2018_RevModPhys90_045006,
rodney_2011_ModellingSimulMatterSciEng19_083001}.
The technical difficulties pertain to the interplay of competing sources of stochasticity --~in particular their self-generated structural disorder~-- and to the resulting out-of-equilibrium nature of these systems.
More generally, these issues are common to driven complex systems in a broad sense, \textit{i.e.}~composed of many interacting degrees of freedom, and stretch to such dissimilar fields as active matter \cite{marchetti_2013_RevModPhys85_1143}
% \cite{marchetti_2013_RevModPhys85_1143} "Hydrodynamics of soft active matter"
or machine learning \cite{baityjesi_2019_JStatMech2019_124013,geiger_2019_PhysRevE100_012115}.
% \cite{baityjesi_2019_JStatMech2019_124013} "Comparing dynamics: deep neural networks versus glassy systems" 
% \cite{geiger_2019_PhysRevE100_012115} "Jamming transition as a paradigm to understand the loss landscape of deep neural networks"
%
Assessing both the similarities and discrepancies in the statistical features of such systems upon different interactions and drivings, is thus key to allow for the transfer of known results between them.
In that respect, the limit of infinite dimension provides a particularly valuable vantage point, as it can often be exactly analytically tractable and thus allow for rigourous analogies already at a mean-field level.

%------

Dense many-body systems of pairwise interacting particles constitute a standard model for amorphous materials, allowing us to focus on the key role of their structural (positional) disorder.
At sufficiently high packing fraction and low temperature, they behave as amorphous \emph{solids},
meaning that they exhibit a rigidity which is sustained under external shear deformation,
up to a sample-dependent maximum shear strain amplitude.
Beyond this so-called `yielding point', they might reach a steady state with a finite built-in stress, and thus behave as yield stress \emph{fluids}.
Such a transition from arrested to flowing state is to be expected in driven disordered systems, and has prompted  the characterisation of corresponding phase diagrams as for instance in Refs~\cite{liu_nagel_1998_Nature396_6706,bi_manning_2016_PhysRevX6_021011}.
%\cite{liu_nagel_1998_Nature396_6706} "Nonlinear dynamics: Jamming is not just cool any more"
%\cite{bi_manning_2016_PhysRevX6_021011} "Motility-Driven Glass and Jamming Transitions in Biological Tissues"
%
However, one important pending question is the following: are global \textit{versus} local drivings fundamentally different? 
This question should obviously be completed with the nature of the observables of interest,
on the one hand mean-field quantities such as pressure or the average stress (taken as a proxy for the predicted macroscopic stress),
and on the other hand spatio-temporal correlations and their related features (such as possible transient or permanent shear bands).
While the latter are \textit{a priori} highly sensitive to the built-in spatio-temporal structure of a specific driving,
mean-field quantities are
better suited --~by their very definition~-- for establishing possible equivalences between different drivings.

%------

For sheared amorphous materials,
the equivalence between global shear strain and random local displacements can be addressed analytically in the limit of infinite spatial dimension, where their statics and dynamics become exactly mean-field.
This limit has been extensively studied in the past years since it provides an exact benchmark for investigating the  properties of structural glasses~\cite{book_parisi_urbani_zamponi_2020}, such as the statistical features of their free-energy landscape~\cite{charbonneau_2014_NatureCommunications5_3725},
their equilibrium behavior~\cite{maimbourg_2016_PhysRevLett116_015902,kurchan_2016_JStatMech2016_033210,szamel_2017_PhysRevLett119_155502} or their response to quasistatic drivings \cite{rainone_urbani_2016_JStatMech2016_053302,biroli_urbani_2016_NatPhys12_1130,
urbani_zamponi_2017_PhysRevLett118_038001,
biroli_urbani_2018_SciPostPhys4_020,altieri_2019_PhysRevE100_032140}.
In fact, this framework can naturally be extended to the new driving protocol that has been recently introduced in Ref.~\cite{morse_roy_agoritsas_2020_Arxiv-2009.07706}, namely the quasistatic driving of a glass through random local displacements, constant on each particle and spatially correlated.
Under this new Athermal Quasistatic Random Displacements (AQRD) protocol, it has been shown in Ref.~~\cite{morse_roy_agoritsas_2020_Arxiv-2009.07706} for two-dimensional numerical simulations (of Hertzian contact particles, under periodic boundary conditions)
that the stress-strain curves are qualitatively similar to those obtained under a standard Athermal Quasistatic Shear (AQS) protocol~\cite{maloney_lemaitre_2006_PhysRevE74_016118},
as well as the distributions in the pre-yielding regime 
of \textit{(i)}~local elastic moduli along elastic branches,
\textit{(ii)}~strain intervals between stress drops,
and \textit{(iii)}~stress drop magnitudes.
More importantly, it was shown that these mean-field-like metrics  can be quantitatively collapsed one onto each other, in remarkably good agreement with the infinite-dimensional predictions.
The aim of the present paper is to present the detailed derivation of the exact mean-field dynamics which led to these predictions, and to discuss its implications for and beyond quasistatics.
Note however that our formalism allows a finite temperature, so our results will not be restricted to the strictly athermal case; thereafter we will keep by convention the abbreviations AQRD and AQS to refer to the two types of quasistatic driving even at finite temperature.
Conceptually speaking, our main statement will be the following:
in the infinite-dimensional limit, traversing the potential energy landscape of such many-body systems is equivalent under a global shear or under a constant random local displacement field,
in the sense that the statistical sampling of the configurational phase space leads to the same mean-field metrics, up to a single rescaling factor. 

%------

Thereafter we essentially adapt the derivation for the case of a global shear strain presented in Ref.~\cite{agoritsas_maimbourg_zamponi_2019_JPhysA52_334001}, following the notations and definitions of the recent extensive review on this topic~\cite{book_parisi_urbani_zamponi_2020}.
We start in Sec.~\ref{section-settings-global-vs-local-shear-strain} by defining the random local displacements settings we consider, mirroring the global shear case,
and we discuss in particular how we choose to encode their spatial correlations.
Secondly we sketch in Sec.~\ref{section-recalling-DMFE} how we can go from the full many-body dynamics to an effective scalar stochastic process with such random local displacements.
Thirdly we focus in Sec.~\ref{section-quasistatics-glassy-states} on the quasistatic driving of a glassy state, starting from a replica-symmetric (RS) equilibrium configuration, and connect with the previously \emph{static} results for quasistatic shear in infinite-dimension.
The latter has first been discussed for hard spheres~\cite{rainone_2015_PhysRevLett114_015701} and its further extensions and ramifications
%\cite{rainone_urbani_2016_JStatMech2016_053302,urbani_zamponi_2017_PhysRevLett118_038001,
%biroli_urbani_2018_SciPostPhys4_020,altieri_2019_PhysRevE100_032140}
are extensively reviewed in Ref.~\cite{book_parisi_urbani_zamponi_2020}.
In Sec.~\ref{section-AQRD-stress-strain-elastic-modulus} we focus furthermore on the implications for the quasistatic stress-strain curves and the elastic modulus.
Finally, in Sec.~\ref{section-discussion-conclusion}, we conclude and discuss some implications as possible perspectives to this work.

%------

This whole derivation will essentially allow us to show how our random local forcing and global shear turn out to be strictly equivalent, in the infinite-dimensional limit, upon a simple rescaling of the accumulated strain: the scaling factor is then simply controlled by the variance of relative local displacements for a given pair of interacting particles, which encodes the finite spatial correlations of the local displacement field.
This statement holds in particular for athermal quasistatic drivings, and the AQS protocol can be interpreted, from that perspective, as a special case of the AQRD protocol.
Note that this statement will be obtained for the replica-symmetric equilibrium case.
Further work would be needed to extend the AQRD protocol to the full-replica-symmetry-breaking case (and thus all the way down to the yielding transition), as it has already been done for shear~\cite{rainone_urbani_2016_JStatMech2016_053302}.

%_____________________________________________________________________________________________________
\section{Global shear \textit{versus} random local displacements}
\label{section-settings-global-vs-local-shear-strain}

We consider the same general settings as in Refs.~\cite{agoritsas_maimbourg_zamponi_2019_JPhysA52_144002,agoritsas_maimbourg_zamponi_2019_JPhysA52_334001}:
a system of $N$ interacting particles in $d$ dimension and of positions ${\lbrace {\bf x}_i(t) \in  \Omega \subset \mathbb{R}^d \rbrace_{i=1,\dots,N}}$ at time $t$.
The region $\Omega$ has a volume ${\vert \Omega \vert}$ and thus a number density ${\rho = N/\vert \Omega \vert}$, and for simplicity we assume $\Omega$ to be a cubic region with periodic boundary condition (\textit{i.e.}~in the same spirit as the numerical settings for instance in Ref.~\cite{morse_roy_agoritsas_2020_Arxiv-2009.07706}).
Note that we always assume first the thermodynamic limit (${N \to \infty}$ and ${\vert \Omega \vert \to \infty}$ at fixed $\rho$), and secondly the infinite-dimensional limit (${d \to \infty}$).
We consider the case of pairwise interactions between identical particles, with a generic radial potential ${v (\vert \rr_{ij}(t) \vert)}$ where ${\rr_{ij}(t) = {\bf x}_{i}(t) - {\bf x}_{j}(t)}$ fluctuates around a typical interaction length~$\ell$.
This potential could be chosen to be a hard-sphere, soft-sphere or Lennard-Jones-like for instance, as long as it is thermodynamically stable in high dimension and has a well-defined infinite-dimensional limit upon rescaling:
${\lim_{d \to \infty} v (\ell (1 + h/d)) = \redv(h)}$ \cite{book_parisi_urbani_zamponi_2020}\footnote{
We recall for instance from Sec.~2.1 of Ref.~\cite{kurchan_2016_JStatMech2016_033210} that we have
%for hard spheres at inverse temperature $\beta$ that ${e^{-\beta v(r)}=\theta(r-\ell) = \theta (h)= e^{-\beta \redv(h)}}$,
for soft harmonic spheres ${v(r)=\epsilon \, d^2 (r/\ell-1)^2 \, \theta(\ell -r) = \epsilon \, h^2 \, \theta (-h)= \redv(h)}$,
for soft spheres ${v(r)= \epsilon \, (\ell/r)^{\alpha d} \stackrel{(d\to\infty)}{\to} \epsilon \, e^{-\alpha h} = \redv(h)}$,
and for Lennard-Jones potential ${v(r) = \epsilon \, \argc{(\ell/r)^{4d} - (\ell/r)^{2d}} \stackrel{(d\to\infty)}{\to} \epsilon \argc{e^{-4h} - e^{-2h}} = \redv(h)}$.
}.
The rationale behind this requirement is that, in the infinite-dimensional limit, the interparticle distances (for effectively interacting particles) have fluctuations of ${\mathcal{O}(1/d)}$ around $\ell$, so that ${r_{ij}(t) = \ell (1 + h_{ij}(t)/d)}$ with the gap ${h_{ij}(t) \sim \mathcal{O}(d^0)}$.
The definition of the rescaled potential ${\redv(h)}$ allows to focus on this gap of order $1$.

A global shear strain of amplitude $\gamma$, in the plane ${\lbrace \hat{\bf x}_1,\hat{\bf x}_2 \rbrace}$ for instance, is defined by
\begin{equation}
 \hat{\gamma} = \left( \begin{array}{cccc}
  0 & \gamma & 0 & \cdots
  \\
  0 & 0 & 0 & \cdots
  \\
  \vdots & \vdots & \vdots
 \end{array} \right)
 \in \mathbb{R}^d \times \mathbb{R}^d
 \quad \Rightarrow \quad
 {\bf x}_i'
 	= \underbrace{\argp{\hat{1} + \hat{\gamma}}}_{= \hat{S}_{\gamma}} {\bf x}_i
 	= \argp{\begin{array}{c} x_{i,1} + \gamma x_{i,2} \\ x_{i,2} \\ \vdots \end{array}}
 	= {\bf x}_i + \gamma \, x_{i,2} \, \hat{\bf x}_1
 	\, ,
\end{equation}
meaning that it assigns to each particle~$i$ a local displacement ${\gamma \cc_i}$ with ${\cc_i= x_{i,2} \, \hat{\bf x}_1}$.
This local displacement is always applied along the direction ${\hat{\bf x}_1}$ and its amplitude depends on the configuration on which the shear strain step is applied.
It is thus continually updated along the dynamics.
The motion of particles in the laboratory frame can then be decomposed into the affine motion due to the accumulated shear strain ${\gamma(t)}$, starting from a given initial configuration, and a `non-affine' correction:
\begin{equation}
 {\bf x}_i(t)
 	= {\bf x}_i(0) + \gamma(t) \, x_{i,2}(0) \, \hat{\bf x}_1 + \uu_i(t)
 \: \Rightarrow \:
 \rr_{ij}(t)
 	= \underbrace{\rr_{ij}(0) + \gamma(t) \, r_{ij,2}(0) \, \hat{\bf x}_1}_{=\rr_{0,ij}'(t) \text{ (for shear)}} + \vv_{ij}(t)
 \, ,
 	%\equiv \rr_{0,ij}'(t) + \vv_{ij}(t)
\label{eq-def-non-affine-shear}
\end{equation}
with ${\uu_i(t)}$ and ${\vv_{ij}(t)}$ the non-affine absolute and relative displacements, respectively.
Note that, since the affine transformation is the same for both the absolute and relative positions (using the matrix ${\hat{S}_{\gamma}}$) it makes sense to define a `co-shearing frame' where coordinates are directly given by the non-affine motion~\cite{maloney_lemaitre_2006_PhysRevE74_016118}.
For random local displacements, we essentially release the constraint for all these vectors ${\lbrace \cc_i \rbrace_{i=1,\dots,N}}$ to be aligned to ${\hat{\bf x}_1}$, and we allow for the local displacement vector ${\vert c \rangle \equiv \lbrace \cc_i \rbrace_{i=1,\dots,N}}$ to be a \emph{constant} random vector in ${\mathbb{R}^{Nd}}$.
We can then generalise the definition of non-affine motion from Eq.~\eqref{eq-def-non-affine-shear} as follows:
\begin{equation}
 {\bf x}_i(t)
 	= {\bf x}_i(0) + \gamma(t) \, \cc_i + \uu_i(t)
 \: \Rightarrow \:
 \rr_{ij}(t)
	= \underbrace{\rr_{ij}(0) + \gamma(t) \, \left( \cc_i -\cc_j \right)}_{=\rr_{0,ij}'(t) \text{ (for random local displacements)}} + \vv_{ij}(t)
 \, ,
	%\equiv \rr_{0,ij}'(t) + \vv_{ij}(t)
\label{eq-def-non-affine-AQRD}
\end{equation}
%where ${\tilde{\gamma}(t)}$ is the `random strain' associated to this driving.
%
Contrary to global shear, we cannot define a `co-shearing frame' \textit{per se}, but we can still focus on the non-affine motion through ${\uu_i(t)}$ and ${\vv_{ij}(t)}$.

The definitions~\eqref{eq-def-non-affine-shear}-\eqref{eq-def-non-affine-AQRD} will be key in the next section, when we focus on obtaining an effective dynamics for the non-affine motion: the derivation will be very similar for shear and random local displacements, and their differences will mostly rely on replacing ${\rr_{0,ij}'(t)}$ by their corresponding explicit expressions.
Moreover, regarding quasistatic driving of glasses (and thus the connection between AQS and AQRD), what will be relevant is specifically the statistical distribution of the affine motion at finite shear strain.
For AQRD, it is thus the distribution of the \emph{relative} local displacement of pairs of particles ${\cc_{ij}=\cc_i -\cc_j}$ that will matter, through the definition of the associated affine motion ${\rr_{0,ij}'(t)}$.
An important point, though, is that we must impose the scaling in distribution ${\vert \cc_{ij} \vert^2 \sim \mathcal{O}(1/d)}$ in order to mirror the scaling of local displacements under global shear.
In the latter case we have indeed ${\vert r_{ij,2}(0)^2 \hat{{\bf x}}_1^2 \vert=r_{ij,2}(0)^2 \sim \mathcal{O}(1/d)}$,
since it involves one component of a $d$-dimensional vector ${\rr_{ij}(0)}$ whose norm is of order $1$.
If we were considering a finite system, we should moreover take care of the finite-size and finite-$N$ scalings, as discussed in Ref.~\cite{morse_roy_agoritsas_2020_Arxiv-2009.07706}.
Here we recall that we consider on the contrary an infinite-size system and the thermodynamic limit, so there is no such issue.

In order to implement a local displacement vector ${\vert \lbrace c_i \rbrace \rangle \in \mathbb{R}^{Nd}}$ as in the AQRD protocol introduced in Ref.~\cite{morse_roy_agoritsas_2020_Arxiv-2009.07706}, we first generate a local displacement field ${\mathcal{C}({\bf x})}$ continuously defined for ${{\bf x} \in \mathbb{R}^d}$, with a Gaussian distribution given by
\begin{equation}
\overline{\mathcal{C}({\bf x})}=0
\, ,\quad
\overline{\mathcal{C}({\bf x}) \cdot \mathcal{C}({\bf x'})}= \ell^2 \Xi \, f_{\xi}(\vert {\bf x}-{\bf x'} \vert)/d
\, .
\end{equation}
The overline denotes the statistical average over this `quenched random displacement field',
$\Xi$ is a tunable amplitude which has the units of a length,
and ${\xi>0}$ a finite correlation length.
For practical purposes, whenever we need an explicit expression we will assume a Gaussian function ${f_{\xi}(x)=e^{-x^2/(2\xi^2)}/\sqrt{2 \pi \xi^2}}$.
Secondly we associate to each particle a local displacement fixed by the value of the displacement field at its initial position.
The local displacements are then also Gaussian distributed:
\begin{equation}
\cc_i = \mathcal{C}({\bf x}_i(0)
\quad \Rightarrow \quad
\overline{\cc_i} = 0 \, , \quad \overline{\cc_i \cdot \cc_j} = \ell^2 \Xi \, f_{\xi}(\vert \rr_{ij} (0) \vert)/d
\, .
\end{equation}
Note that the explicit scaling in $d$ is chosen precisely such as %${ \ell^2 \Xi \sim \mathcal{O}(1)}$
to match the scaling of local displacements in AQS, as discussed above.
Moreover, we implicitly assume a statistical isotropy of the local displacements by considering a \emph{radial} function for the correlator.
From there, we can directly characterise the distribution of the relative local displacements
${\cc_{ij}=\cc_i -\cc_j}$.
It is also Gaussian of zero mean,
but its variance considerably simplifies in the infinite-dimensional limit:
using that ${r = \ell (1 + h/d)}$ with the gap ${h \sim \mathcal{O}(1)}$, we thus have
${f_{\xi} \argp{r_{ij}(0)} = f_{\xi}(\ell) + f_{\xi}'(\ell) \frac{\ell}{d} h_{ij}(0) + \mathcal{O}(1/d^2)}$.
This implies on the one hand that, for different pairs ${(i,j) \neq (i',j')}$, correlations are subdominant in high dimension and eventually vanish in the limit ${d \to \infty}$:
\begin{equation}
\begin{split}
 d \, \overline{\cc_{ij} \cdot \cc_{i'j'}}
 &= \ell^2 \Xi \argc{
	f_{\xi} \argp{r_{ii'}(0)} + f_{\xi}\argp{r_{jj'}(0)} - f_{\xi}\argp{r_{ij'}(0)} - f_{\xi}\argp{r_{i'j}(0)}
 }
 \\
 &= \ell^2 \Xi \, f_{\xi}' (\ell) \frac{\ell}{d} \argc{ h_{ii'}(0) + h_{jj'}(0) - h_{ij'}(0) - h_{i'j}(0)}
 \stackrel{(d \to \infty)}{\to} 0 \, ,
\end{split}
\label{eq-distr-cij-1}
\end{equation}
whereas on the other hand the variance of a given pair remains finite:
\begin{equation}
 d \, \overline{\cc_{ij}^2}
 = 2 \ell^2 \Xi \, \argc{f_{\xi}\argp{0} - f_{\xi}\argp{r_{ij}(0)}}
 \stackrel{(d \to \infty)}{\to} 2 \ell^2 \Xi \, \argc{f_{\xi}\argp{0} - f_{\xi}\argp{\ell}}
 \equiv \mathfrak{F} \argp{\Xi ,\ell, \xi}
 \, .
\label{eq-distr-cij-2}
\end{equation}
Another way to phrase these results is that spatial correlations between the local displacements ${\cc_i = \mathcal{C}({\bf x}_i(0)) }$ only affect the variance of a given pair \emph{relative} displacements ${\cc_{ij}}$.
This is consistent with the fact that different pairs of particles effectively do not interact in infinite dimension (or rather their contribution becomes irrelevant, in the limit ${d \to \infty}$, to path-integral statistical averages).
We can choose to decompose these relative displacements as ${\sqrt{d} \, \cc_{ij} \equiv \tilde{c}_{ij} \, \hat{\cc}_{ij}}$,
where ${\hat{\cc}_{ij}}$ is a unitary vector with a uniform distribution by statistical isotropy.
The factor $\mathfrak{F}$ defined in Eq.~\eqref{eq-distr-cij-2} should then be understood as the variance of the amplitude ${\tilde{c}_{ij}}$, restricted to the pairs whose interactions are relevant for the dynamics (\textit{i.e.}~${r_{ij}(0) \approx \ell}$).
Beware that ${\tilde{c}_{ij}}$ is not the norm, since we allow it to take negative values for a given choice of unitary vector.
Its probability distribution is simply the Gaussian function:
\begin{equation}
 \bar{\mathcal{P}} \argp{\tilde{c}_{ij}}
 	=e^{-\tilde{c}_{ij}^2/(2\mathfrak{F})}/\sqrt{2 \pi \mathfrak{F}}
 \quad \Rightarrow \quad
 \overline{\tilde{c}_{ij}}=0 \, , \quad \overline{\tilde{c}_{ij}^2}=\mathfrak{F}
 \, .
\label{eq-def-frakF-Gaussian-PDF}
\end{equation}
Note that we choose a slightly different convention than in Ref.~\cite{morse_roy_agoritsas_2020_Arxiv-2009.07706}:
here the vector ${\vert c \rangle}$ is a displacement field and has thus the units of a length, and it differs only from a factor $\ell$ compared to the unitless strain field in Ref.~\cite{morse_roy_agoritsas_2020_Arxiv-2009.07706}; otherwise other quantities are defined with the same units, and in particular $\sqrt{\mathfrak{F}}$ has the units of a length.

This dependence of the variance ${\mathfrak{F}}$ on the spatial correlations of local displacements is in fact physically expected and can be simply understood in the following two opposite limits for $\xi$.
For an infinite correlation length $\xi$,
all the ${\lbrace \cc_i \rbrace}$ would be the same, so the whole system is simply translated and there would be no relative displacements; so for ${\xi \to \infty}$ their distribution should consistently tend to ${\mathcal{P}({\bf c_{ij}}) \propto \delta(\vert {\bf c_{ij}} \vert)}$, \textit{i.e.}~ ${\mathfrak{F}=0}$.
In the opposite limit, the variance ${\mathfrak{F}}$ diverges, so we need to have a finite correlation ${\xi>0}$ to keep it regular; it is physically the case when we consider discrete interacting particles instead of a true continuum of local displacements.
If we assume ${f_\xi(x)}$ to be a normalised Gaussian function, we can compute explicitly ${\mathfrak{F}}$:
%and in particular its dependence on what remains encoded of spatial correlation through the length $\xi$:
\begin{equation}
f_{\xi}(x)=\frac{e^{-x^2/(2\xi^2)}}{\sqrt{2 \pi \xi^2}}
\quad \Rightarrow \quad
 \mathfrak{F} \argp{\Xi,\ell, \xi}
 \stackrel{\eqref{eq-distr-cij-2}}{=} \frac{2}{\sqrt{2 \pi}} \ell \Xi \, \frac{\ell}{\xi} \argp{1 - e^{-(\ell/\xi)^2/2}} 
 \left\lbrace \begin{array}{l}
 	\stackrel{(\ell/\xi \gg 1)}{=} \frac{2}{\sqrt{2\pi}}\ell \Xi \, \frac{\ell}{\xi}
 	%= \frac{2}{\sqrt{2\pi}} \ell^2 \Xi \, \frac{1}{\xi}
 	\\
 	\stackrel{(\ell/\xi \ll 1)}{=} \frac{1}{\sqrt{2\pi}} \ell \Xi \, \argp{\ell/\xi}^3
 	%= \frac{1}{\sqrt{2\pi}} \ell^2 \Xi \, \frac{\ell^2}{\xi^3}
 \, .
 \end{array} \right.
\label{eq-mathfrak-F-explicit-Gaussian_fxi}
\end{equation}
We thus predict a crossover of the $\xi$-dependence depending on the ratio ${\ell/\xi}$,
with ${\mathfrak{F} \sim 1/\xi}$ at ${\ell/\xi \gg 1}$
and ${\mathfrak{F} \sim 1/\xi^{3}}$ at ${\ell/\xi \ll 1}$.
Technically, these specific scalings rely on the functional form of the even correlator ${f_{\xi}(x) = \xi^{-1} f_1(x/\xi)}$, which holds in particular for $f_{\xi}$ being a Gaussian function.
Nevertheless, the decrease of ${\mathfrak{F}}$ with an increasing correlation length $\xi$ must be qualitatively robust with respect to alternative (but still physical) choices for ${f_{\xi}(x)}$.

As a concluding remark, we note that such a correlation length exists under global shear (thus for the AQS protocol as well).
I was in fact the initial motivation for the creation of the AQRD protocol in Ref.~\cite{morse_roy_agoritsas_2020_Arxiv-2009.07706}.
Indeed, when we shear a finite-size system with periodic boundary conditions, the individual affine motions are \emph{by construction} correlated on a finite portion of the whole system, with a periodicity of twice the system size (\textit{i.e.}~the value of ${\xi}$ is set once and for all).
Note that this statement is only true in the laboratory frame, since in the co-shearing frame the affine motions are by definition set to zero; the associated local displacements have to be defined in the laboratory  frame in order to be able to establish a direct connection between the global shear and random-local-displacements protocols.
An intermediate case is to have regularly patterned local displacements
(chosen such as to be compatible with the periodic boundary conditions), where we can tune the correlation length $\xi$.
Such settings has also been examined, for two-dimensional numerical simulations, in Ref.~\cite{morse_roy_agoritsas_2020_Arxiv-2009.07706}.
Here we considered the case of Gaussian random local displacements because it directly relates to the shear case in infinite dimension, as we will see in the next section.
However, these different alternatives will only affect the quantitative distribution of affine motions ${\rr_{ij}'(t)}$ but not the effective dynamics for the non-affine motions ${\uu_i(t)}$ and ${\vv_{ij}(t)}$.
So we could \textit{a priori} generalise our derivation to the patterned local displacements as well (but this is kept for future work).

%_____________________________________________________________________________________________________
\section{From many-body to effective scalar dynamics}
\label{section-recalling-DMFE}

We consider the following many-body Langevin dynamics, adapted from the standard shear case~\cite{agoritsas_maimbourg_zamponi_2019_JPhysA52_334001}.
For a finite `shear rate' ${\dot{\gamma}(t)}$ 
(such that the AQRD case corresponds to the limit ${\dot{\gamma}(t) \to 0}$)
we have:
\begin{equation}
 \zeta \argc{\dot{\bf x}_i(t)- \dot{\gamma}(t) \, \cc_i }
 	=  {\bm F}_i(t) %\underbrace{-\sum_{j (\neq i)} \nabla v \argp{{\bf x}_i(t)-{\bf x}_j(t)}}_{\text{interaction with other particles} \, {\bm F}_i(t)} 
 	+ \bm{\xi}_i(t)
 \, , \quad \text{with} \quad
 {\bm F}_i(t) = -\sum_{j (\neq i)} \nabla v \argp{\vert {\bf x}_i(t)-{\bf x}_j(t) \vert}
\label{eqC3:GENLang-shear-dynamics-AQRD}
\end{equation}
and a microscopic Gaussian noise ${\lbrace {\bm \xi}_i(t) \rbrace_{i=1,\dots,N}}$, of mean and variance respectively given by
\begin{equation}
 	\moy{\xi_{i,\mu}(t)}_{\bm\xi} =0 , \qquad
 	\moy{\xi_{i,\mu}(t) \xi_{j,\nu}(t')}_{\bm \xi}
		= \delta_{ij}  \delta_{\mu\nu} [2 T \zeta  \delta(t-t') + \Gamma_C(t,t')]
 \, ,
\label{eqC3:GENLang-shear-noise}
\end{equation}
where the brackets ${\moy{\cdots}_{\bm\xi}}$ denote the statistical average over realisations of the noise~${\bm\xi}$.
$\zeta$ is the (local) friction coefficient and ${T=\beta^{-1}}$ the temperature of a thermal bath (with the Boltzmann constant ${k_B=1}$ setting the units).
The generic noise kernel ${\Gamma_C(t,s)}$ can be chosen such as to describe a wide variety of physical situations, as discussed extensively in Sec.~2 of Ref.~\cite{agoritsas_maimbourg_zamponi_2019_JPhysA52_144002}.
By essentially tuning its `persistence time' $\tau$, we can consider anything
from an isotropic active matter with ${\Gamma_C(t,s) = f_0^2 \, e^{- \valabs{t-s}/\tau}}$,
to a pure thermal bath (${\tau=0}$) with ${\Gamma_C(t,s)=0}$,
and constant random forces (`${\tau=\infty}$') with ${\Gamma_C(t,s)=f_0^2}$.
In addition, we need to specify the distribution of the initial configurations at time ${t=0}$; for explicit computations we can focus on the particular case of an equilibrium (possibly supercooled) liquid phase at a temperature ${T_0=\beta_0^{-1}}$, where positions are sampled from a Gibbs-Boltzmann distribution ${\propto e^{-\frac12 \sum_{i \neq j} v(\vert r_{ij}(0) \vert) }}$.

These settings are very similar to the standard case under shear strain, that we have examined in Ref.~\cite{agoritsas_maimbourg_zamponi_2019_JPhysA52_334001}.
Here we essentially replaced for each particle~$i$ the time-dependent local displacements ${\gamma(t) x_{i,2}(t) \hat{\bf x}_1}$ under a global shear (in the plane ${\lbrace \hat{\bf x}_1 ,\hat{\bf x}_2 \rbrace}$) by the random local displacements ${{\gamma}(t) \, \cc_i}$.
In Ref.~\cite{agoritsas_maimbourg_zamponi_2019_JPhysA52_334001} we focused on the non-affine motion by using the change of variables of Eq.~\eqref{eq-def-non-affine-shear}, which is equivalent to working in the co-shearing frame.
We can similarly use Eq.~\eqref{eqC3:GENLang-shear-dynamics-AQRD} to obtain the dynamics for the non-affine motion under random local displacements:
\begin{equation}
 \zeta \dot{\uu}_{i}(t)
   = {\bm F}_i(t) + \bm{\xi}_i(t)
  \, , \quad \text{with} \quad
 {\bm F}_i(t) = -\sum_{j (\neq i)} \nabla v \argp{\vert \rr_{0,ij}'(t) + \uu_{i}(t) - \uu_{j}(t)) \vert}
 \label{eqC3:GENLang-for-ucs-bis-AQRD}
\end{equation}
with the uniform initial condition ${\uu_i(0)=\bm{0} \, \forall i}$.
Compared to global shear, we do not have a term ${\zeta \hat{\dot{\gamma}}(t) \uu(t)}$ so the dynamics is \textit{de facto} isotropic.
Moreover, the affine motion is given by ${\rr_{0,ij}'(t) \equiv \rr_{0,ij} + \gamma(t) \cc_{ij}}$ instead of ${\rr_{0,ij}'(t) = \hat{S}_{\gamma}(t) \, \rr_{0,ij}}$ (see Eqs.~\eqref{eq-def-non-affine-shear}-\eqref{eq-def-non-affine-AQRD}).
Note that if we start from a statistically isotropic initial condition, such as equilibrium,
with such an isotropic dynamics we can always assume statistical isotropy to hold.

This many-body dynamics becomes exactly mean-field in infinite dimension, as shown in previous works both for the isotropic case~\cite{agoritsas_maimbourg_zamponi_2019_JPhysA52_144002} and in presence of global shear~\cite{agoritsas_maimbourg_zamponi_2019_JPhysA52_334001}.
In a nutshell, this can be understood as a consequence of two key physical assumptions:
\textit{(i)}~particles always stay `close' to their affine motion with respect to their initial position
--~in the sense that their non-affine displacements are of ${\mathcal{O}(1/d)}$~--
and \textit{(ii)}~each particle has numerous uncorrelated neighbours.
% \cite{kirkpatrick_wolynes_1987_PhysRevA_35_3072} "Connections between some kinetic and equilibrium theories of the glass transition" % as cited in Ref.~\cite{kurchan_2016_JStatMech2016_033210}
%
Intuition can be gained from the simpler case of an isotropic random walk: each particle has so many directions towards which it can move, that it effectively explores a volume whose typical radius shrinks with an increasing dimensionality.
Moreover, in a dense system each particle has ${\mathcal{O}(d)}$ neighbours, and once it interacts with a given neighbour it is very unlikely that it would interact with it again, making them effectively uncorrelated.
The recipe for obtaining the mean-field dynamics is thus as follows:
we start from the many-body dynamics~\eqref{eqC3:GENLang-shear-dynamics-AQRD},
focus on the non-affine displacement in Eq.~\eqref{eqC3:GENLang-for-ucs-bis-AQRD},
and Taylor-expand the interaction force ${{\bm F}_i(t)}$ to leading order in the infinite-dimension limit.
Technically, the difficulty is to correctly identify the scaling in $d$ of the different contributions to the dynamics and the leading-order terms.
Physically that amounts to neglecting the contributions due to collective feedbacks involving more than two particles, which are subdominant in the infinite-dimension limit when computing statistical averages of observables.

Starting from Eq.~\eqref{eqC3:GENLang-for-ucs-bis-AQRD}, we can directly adapt the mean-field dynamics previously obtained for global shear (specifically Eqs.~(23)-(24) of Ref.~\cite{agoritsas_maimbourg_zamponi_2019_JPhysA52_334001}).
The interactions with other particles are fully described by three tensorial kernels ${\lbrace  \hat{k}(t), \hat{M}_R(t,s),\hat{M}_C(t,s) \rbrace}$.
As a result of the Taylor-expansion of ${{\bm F}_i(t)}$, these kernels are defined as averages involving the force ${\nabla v(\rr_{ij}(t))=\nabla v(\rr_{0,ij}'(t) + \vv_{ij}(t))}$.
From this point, we can drop on the indices ${(i,j)}$ only keeping in mind that $\cc$ stands for the \emph{relative} local displacements $\cc_{ij}$.
${\rr_0'(t)=\rr_0 + \gamma(t) \, \cc \,}$ is set by the distributions of the initial condition and of the local strains.
As for the non-affine displacements, their dynamics is exactly mean-field and given by the following \emph{vectorial} stochastic processes:
\begin{equation}
\label{eq-vectorial-effective-stoch-processes}
\begin{split}
  & \zeta \dot{\uu}(t)
  = - \hat{k}(t) \, \uu(t)
 	+ \int_0^t \de s \, \hat{M}_R(t,s) \, \uu(s)
 	+ \sqrt{2}\,\bm{\Xi}(t)
 \, ,
 \\
  & \frac{\zeta}{2} \dot \vv(t)
  =	- \frac12 \hat{k}(t) \, \vv(t)
 	+ \frac12 \int_0^t \de s \, \hat{M}_R(t,s) \, \vv(s)
 	- \nabla v \argp{\rr_{0}'(t) + \vv(t)}
 	+ \bm{\Xi}(t)
 \, ,
 \\
 & \uu(0)=0 \, , \quad \vv(0)=0 \, ,
 \\
 &	\moy{\Xi_{\mu}(t)}_{\bm{\Xi}}=0 \, , \quad
	\moy{\Xi_{\mu}(t) \Xi_{\nu}(t')}_{\bm{\Xi}}
	= \delta_{\mu\nu} \argc{ T \zeta \delta(t-t') +\frac12 \Gamma_C(t,t') } + \frac12 M^{\mu\nu}_C(t,t')
 \, ,
\end{split}
\end{equation}
where the noise has a zero mean by statistical isotropy,
and the following self-consistent equations for the tensorial kernels:
\begin{equation}
\label{eqC3:Mself}
\begin{split}
 k^{\mu\nu}(t)
 	&= \rho \int \de \rr_0 \, g_{\text{in}}(\rr_0) \int \de \cc \, \bar{\mathcal{P}}(\cc) \,
 		\moy{ \nabla_\mu \nabla_\nu v( \rr_0'(t) + \vv(t))}_{\vv}
 \ , \\
 M^{\mu\nu}_C(t,t')
 	&=  \rho \int \de \rr_0 \, g_{\text{in}}(\rr_0) \, \int \de \cc \, \bar{\mathcal{P}}(\cc) \, 
 		\moy{ \nabla_\mu v(\rr_0'(t) + \vv(t))  \nabla_\nu v(\rr_0'(t') + \vv(t')) }_{\vv}  
 \ , \\
 M^{\mu\nu}_R(t,s)
 	&= \rho  \int \de \rr_0 \, g_{\text{in}}(\rr_0)  \int \de \cc \, \bar{\mathcal{P}}(\cc) \, 
 		\, \left. \frac{\delta \moy{  \nabla_\mu v(\rr_0'(t) + \vv(t)) }_{\vv,\bm{P}}}{\delta P_{\nu}(s)}\right\vert_{\bm P=\bm0}
 \ .
\end{split}
\end{equation}
Here $\rho$ is the density, 
${g_{\text{in}}(\rr_0)}$ the initial distribution of inter-particle distances in the laboratory frame at ${t=0}$~\cite{agoritsas_maimbourg_zamponi_2019_JPhysA52_144002,book_hansen_mcdonald-4th-ed},
and ${\bar{\mathcal{P}}(\cc)}$ the distribution of relative local displacements.
The brackets ${\moy{\bullet}_{\vv}}$ denote the dynamical average over the stochastic process $\vv(t)$ (thus self-consistently defined), starting from a given set ${\lbrace \rr_0, \cc \rbrace }$.
${\bm{P}(s)}$ is to be understood as a perturbation field added into the dynamics, inside the interaction potential as ${\nabla v(\rr_0'(t)+\vv(t) - \bm{P}(t))}$.
If we start from a replica-symmetric equilibrium at inverse temperature ${\beta_0}$, we would have ${g_{\text{in}}(\rr_0)=g_{\rm{eq}}(\rr_0)=e^{-\beta_0 \, v( \rr_0 )}}$.
Finally, as discussed extensively in the previous section around Eq.~\eqref{eq-distr-cij-2}, we consider specifically vectors ${\cc \equiv \tilde{c} \, \hat{\cc} / \sqrt{d} }$ with ${\hat{\cc}}$ a unitary vector uniformly distributed (in order to enforce statistical isotropy).
The norm ${\tilde{c} \sim \mathcal{O}(1)}$ has a Gaussian distribution of zero mean and variance ${\mathfrak{F}}$ given in Eq.~\eqref{eq-def-frakF-Gaussian-PDF}.
Physically, we emphasise that
\textit{(i)}~$\hat{k}(t)$ characterises the averaged divergence of forces at time $t$,
\textit{(ii)}~${\hat{M}_C(t,t')}$ is nothing but the two-time force correlator,
and \textit{(iii)}~${\hat{M}_R(t,s)}$ is the averaged response on the force at time $t$ after a perturbation at an anterior time $s$.

We can in fact reduce further these high-dimensional vectorial stochastic processes.
For that purpose, we use that the pairwise interaction potential is a radial function of the distance
${r(t) \equiv \vert \rr (t) \vert = \vert {\rr_0'(t) + \vv(t)} \vert \approx \ell (1 + h(t)/d)}$,
where $\ell$ is the typical interaction length and the gap ${h(t)\sim \mathcal{O}(1)}$.
This definition of the gap allows us to focus on fluctuations of $\mathcal{O}(1/d)$ around ${r(t) \approx \ell}$.
Exactly as for shear (in Sec.~IV.B of Ref.~\cite{agoritsas_maimbourg_zamponi_2019_JPhysA52_334001}), we can decompose this inter-particle distance in three contributions as follows:
\begin{equation}
\label{eq-def-fluctuating-gap}
 \valabs{\rr(t)}
 	\equiv \valabs{\rr_0'(t) + \vv(t)}
 	\approx r_0'(t) + \hat{\rr}_0'(t) \cdot \vv(t) + \frac{\vv(t)^2 }{2 r_0'(t)}
 	\quad \text{with} \quad \left\lbrace \begin{array}{l}
 		\hat{\rr}_0'(t) \cdot \vv(t) \equiv \frac{\ell}{d} y(t) \, ,
 		\\ \\
 		\frac{\vv(t)^2 }{2 r_0'(t)}
 			\approx \frac{\moy{\vv(t)^2} }{2 \ell}
 			\approx \frac{\ell}{d} \Delta_r(t) \, .
 	\end{array} \right.
\end{equation}
To get to this decomposition,
first we used that ${\valabs{\vv(t)} \ll r_0'(t)}$ in high dimension,
then we defined ${y(t)}$ as the projection of ${\vv(t)}$ on the affine relative displacement ${\hat{\rr}_0'(t)}$,
and finally we introduced the rescaled mean-square-displacement (MSD) function ${\Delta_r(t) = \frac{d}{\ell^2} \moy{\valabs{\uu(t)}^2}}$.
We recall that for a global shear we would have instead ${\rr_0'(t)=\rr_0 + \gamma(t) \,  r_{0,2} \, \hat{\bf x}_1}$, so the relative displacements $\cc$ would be fully prescribed by the initial condition as ${\cc = r_{0,2} \hat{\xx}_1}$.
This corresponds equivalently to the special case where ${\bar{\mathcal{P}}(\cc) = \delta (\cc - r_{0,2} \hat{\xx}_1)}$.
The implications of the more generic definition ${\rr_0'(t)=\rr_0 + \gamma(t) \, \cc}$ are better reflected on the gap associated to the third contribution in Eq.~\eqref{eq-def-fluctuating-gap}, namely the affine contribution ${r_0'(t)}$:
\begin{equation}
 r_0'(t)
 	= \left\vert \rr_{0}+ \gamma(t) \, \frac{\tilde{c}}{\sqrt{d}} \, \hat{\cc} \right\vert
 	\stackrel{(d \to \infty)}{\approx}
 		r_0 \argp{1 + \frac{h_0}{d} + \frac{\gamma(t)}{d} \frac{\tilde{c}}{\, r_0} \, \sqrt{d} \hat{\rr}_{0} \cdot \hat\cc + \frac{\gamma(t)^2}{2d} \frac{\tilde{c}^2}{r_0^2}}
 \, ,
\label{eq-drift-r0-AQRD}
\end{equation}
where $h_0$ is the initial gap, and we truncated higher orders in ${1/d}$.
We recall that ${r_0 \approx \ell}$.
Note that the scalar product of the two random unit vectors ${\hat{\rr}_{0} \cdot \hat\cc}$ scales in distribution as ${1/\sqrt{d}}$, despite the fact that it is a sum of $d$ terms each of ${\mathcal{O}(1/d)}$: this is due to the randomness of both vectors.
We have thus ${g_c \equiv \sqrt{d} \hat{\rr}_{0} \cdot \hat\cc \sim \mathcal{O}(1)}$, normal distributed with zero mean and unit variance.
Further, we can rewrite Eqs.~\eqref{eq-def-fluctuating-gap}-\eqref{eq-drift-r0-AQRD} for the total gap as
\begin{equation}
\label{eq-def-gap-in-AQRD}
 	\valabs{\rr(t)} = \valabs{\rr_0'(t) + \vv(t)}
 	\approx \ell (1 + h(t)/d)
 \quad \text{with} \quad \left\lbrace \begin{array}{l}
 	h(t) = h_0'(t) + y(t) + \Delta_r(t) \, ,
 	\\ \\
 	h_0'(t)= h_0 + \gamma(t) \, g_c \, (\tilde{c}/\ell) + \frac{\gamma(t)^2}{2} \argp{\tilde{c}/\ell}^2 \, .
 	\end{array} \right.
\end{equation}
This allows us to rewrite the high-dimensional \emph{vectorial} stochastic processes Eqs.~\eqref{eq-vectorial-effective-stoch-processes}-\eqref{eqC3:Mself} into a \emph{scalar} stochastic process, with its three associated kernels (generalizing Eqs.~(46) and~(48) from Ref.~\cite{agoritsas_maimbourg_zamponi_2019_JPhysA52_334001}):
\begin{equation}
 \label{eq-def-longitudinal-motion-stoch-process-AQRD}
\begin{split}
 & \widehat{\zeta} \dot y(t)
	= 	- \kappa^{\text{iso}}(t)  y(t)
 		+ \int_0^t \de s \, \mathcal{M}^{\text{iso}}_R(t,s) \, y(s)
 		- \redv'(h(t))
 		+ \Xi(t)
 \, ,
 \\
 & 	h(t)
	 =	h_0 + \gamma(t) \, g_c \, \breve{c} + \frac{\gamma(t)^2}{2} \breve{c}^2
	 		+ y(t) + \Delta_r(t)
 \, ,
 \\
 & \text{Initial condition:}
 	\quad y(0)=0 \, , \quad \gamma(0)=0 \, , \quad \Delta_r(0)=0 \, ,
 	\quad \text{random} \: \arga{h_0, \breve{c}=\tilde{c}/\ell, g_c} \sim \mathcal{O}(1)
 \, ,
 \\
 & \text{Gaussian noise:} \quad \moy{\Xi(t)}_\Xi=0 \, ,  \quad \moy{\Xi(t)\Xi(s)}_\Xi= 2T \hat{\zeta} \delta(t-s) + \mathcal{G}_C(t,s)+ \mathcal{M}^{\text{iso}}_C(t,s)
 \, ,
\end{split}
\end{equation}
with the rescaled friction coefficient ${\hat{\zeta}= \frac{\ell^2}{2 d^2} \zeta}$ and noise kernel  ${\mathcal{G}_C(t,t')= \frac{\ell^2}{2 d^2} \Gamma_C (t,t')}$.
The three rescaled kernels are then:
\begin{equation}
\label{eq-def-kernels-high-dim-gap-AQRD}
\begin{split}
 \kappa^{\text{iso}}(t)
 	&= \frac{\widehat{\varphi}}2  \int^{\infty}_{-\infty}\!\!  \de h_0 \, e^{h_0} g_{\text{in}}(h_0) \, \int \de \tilde{c} \int \de g_c \, \bar{\mathcal{P}}(g_c,\tilde{c}) \, \moy{ \redv''(h(t)) + \redv'(h(t)) }_{h \vert h_0,\tilde{c},g_c}
 \ , \\
 \mathcal{M}_C^{\text{iso}}(t,t')
 	&=  \frac{\widehat{\varphi}}2 \int^{\infty}_{-\infty}\!\!  \de h_0 \, e^{h_0} g_{\text{in}}(h_0) \, \int \de \tilde{c} \int \de g_c \, \bar{\mathcal{P}}(g_c,\tilde{c}) \,
 		\moy{ \redv'(h(t)) \redv'(h(t')) }_{h \vert h_0,\tilde{c},g_c}
 \ , \\
 \mathcal{M}_R^{\text{iso}}(t,s)
 	&=  \frac{\widehat{\varphi}}2 \int^{\infty}_{-\infty}\!\!  \de h_0 \, e^{h_0} g_{\text{in}}(h_0) \, \int \de \tilde{c} \int \de g_c \, \bar{\mathcal{P}}(g_c,\tilde{c}) \,
		\left. \frac{\delta \moy{ \redv'(h(t))  }_{h \vert h_0,\tilde{c},g_c,\mathcal{P}}}{\delta \mathcal{P}(s)}\right\vert_{\mathcal{P}=0}
 \ .
\end{split}
\end{equation}
We recall from Refs.~\cite{agoritsas_maimbourg_zamponi_2019_JPhysA52_144002}-\cite{agoritsas_maimbourg_zamponi_2019_JPhysA52_334001} that instead of the density $\rho$ we have in those definitions a dependence on the rescaled packing fraction ${\widehat{\varphi} = \rho V_d \ell^d/d}$,
with ${V_d = \pi^{d/2} / \Gamma(d/2+1)}$ the volume of the unit sphere in $d$ dimensions.
If we start from equilibrium at temperature ${T_0=\beta_0^{-1}}$, we simply have
${g_{\text{in}}(h_0) = g_{\text{eq}}(h_0) = e^{-\beta_0 \redv(h_0)}}$.
The statistical average over the relative local displacements takes then the explicit form
\begin{equation}
 \int \de \tilde{c} \int \de g_c \, \bar{\mathcal{P}}(g_c,\tilde{c}) \, (\dots)
 \quad \mapsto \quad
	\int_{\mathbb{R}} \underbrace{\de g_c \frac{e^{-g_c^2/2}}{\sqrt{2\pi}}}_{\equiv \mathcal{D} g_c}
	\, \int_{\mathbb{R}} \de \tilde{c} \, \frac{e^{-\tilde{c}^2/(2\mathfrak{F})}}{\sqrt{2 \pi \mathfrak{F}}}
 	 \, (\dots)
 	 \, .
\label{eq-def-P-gc-ctilde}
\end{equation}
Finally, the dynamical equations for the correlation and response functions (and in particular of the MSD function) are strictly the same as for the case of global shear, see Ref.~\cite{agoritsas_maimbourg_zamponi_2019_JPhysA52_334001} (specifically Eq.~(41) and the definitions Eqs.~(31)-(33) and~(40)).
If we wish to add an acceleration term with a particle mass $m$ and/or a retarded friction kernel ${\Gamma_R(t,s)}$
--~in the full many-body dynamics~\eqref{eqC3:GENLang-shear-dynamics-AQRD} that we took as a starting point~-- we can similarly adapt the dynamical equations given in Ref.~\cite{agoritsas_maimbourg_zamponi_2019_JPhysA52_144002} (see specifically its summary section~VII).

We emphasise that the case of global shear can be seen as the special case of Eqs.~\eqref{eq-drift-r0-AQRD}-\eqref{eq-def-gap-in-AQRD}, with
\begin{equation}
 \rr_0'(t) = \rr_0 + \gamma(t) \, r_{0,2} \hat{\xx}_1
 \quad \Rightarrow \quad
 \cc = r_{0,2} \hat{\xx}_1
 \: \Leftrightarrow \:
 \left\lbrace \begin{array}{ll}
 	\text{rescaled amplitude:} & \tilde{c}/r_0 \equiv \sqrt{d} \, c/r_0 = \sqrt{d} \, \hat{r}_{0,2} \equiv g_2
 	\\  
 	\text{direction:} & \hat{\cc}=\hat{\xx}_1 \; \Rightarrow \; g_c = \sqrt{d} \, \hat{r}_{0,1} \equiv g_1 
 \end{array} \right.
% \: \Rightarrow \:
% \int \de \tilde{c} \int \de g_c \, \bar{\mathcal{P}}(g_c,\tilde{c}) \, (\dots)
% = \int \mathcal{D} g_1 \int \mathcal{D} g_2 \, (\dots)
\label{eq-translation-global-local-shear}
\end{equation}
where we used the definition for the unit vector coordinates
${\hat{r}_{0,\mu}\equiv g_\mu/\sqrt{d}}$ with ${g_\mu \sim \mathcal{O}(1)}$~\cite{biroli_urbani_2018_SciPostPhys4_020,agoritsas_maimbourg_zamponi_2019_JPhysA52_334001}.
The average over relative displacements~\eqref{eq-def-P-gc-ctilde} then simplifies into ${\int \mathcal{D} g_1 \int \mathcal{D} g_2 \, (\dots)}$.
In other words, \emph{global shear has the same scalar mean-field dynamics as random local displacements with ${\mathfrak{F}/\ell^2 \equiv 1}$}.
The other way around, 
we can tune the variance ${\mathfrak{F}}$, for instance by playing with the amplitude~$\Xi$ or the spatial correlation length~$\xi$ in Eq.~\eqref{eq-distr-cij-2}, allowing for a more general family of such random local driving.
In the expression~\eqref{eq-def-longitudinal-motion-stoch-process-AQRD} of the gap ${h_0'(t)}$,
if we define ${\gamma_{\text{eff}}(t)= \gamma(t) \, \sqrt{\mathfrak{F}}/\ell}$,
then $\breve{c}$ can be taken Gaussian distributed of zero mean and unit variance, exactly as the random variable ${g_2}$ appearing in the global shear expressions.
This means that we can go one step further than Eq.~\eqref{eq-def-P-gc-ctilde}:
\begin{equation}
 \int \de \tilde{c} \int \de g_c \, \bar{\mathcal{P}}(g_c,\tilde{c}) \, \argp{\dots \arga{\gamma(t), \tilde{c}, g_c} \dots}
 \quad \mapsto \quad
	\int \mathcal{D} g_c \, \int \mathcal{D} \breve{c} \, \argp{\dots \arga{\gamma_{\text{eff}}(t)=\gamma(t) \, \sqrt{\mathfrak{F}}/\ell , \ell \breve{c}, g_c} \dots}
 	 \, .
\label{eq-def-P-gc-ctilde-bis}
\end{equation}
Consequently, in the limit of infinite dimension, the many-body dynamics under either a global shear or spatially correlated random local displacements can be exactly reduced on scalar mean-field dynamics which are strictly equivalent, upon the above rescaling of the accumulated strain.
This is our main result.
Furthermore, the dependence on the spatial correlation length~$\xi$ persists through the variance ${\mathfrak{F}}$ of relative local displacements.
The functional form of ${\mathfrak{F}}$ is for instance given by Eq.~\eqref{eq-mathfrak-F-explicit-Gaussian_fxi} for Gaussian distributed local displacements.

Technically, this simple connection with global shear relies ultimately on the separation of the affine and non-affine motions ${\rr_0'(t)}$ and ${\vv (t)}$, respectively.
In infinite dimension, the explicit dependence on the `accumulated shear strain' ${\gamma(t)}$ only appears within the affine motion, which keeps a memory of the initial condition ${\rr_0'(t)}$ and its associated gap ${h_0'(t)}$.
As for the non-affine motion,
the feedback of $\gamma(t)$ %on the non-affine motion
is treated in an exact mean-field way through the three kernels~\eqref{eq-def-kernels-high-dim-gap-AQRD}.
Our new protocol essentially allows for a more general distribution of relative displacement amplitude ${\bar{\mathcal{P}}(\tilde{c})}$, admittedly Gaussian with zero mean and variance ${\mathfrak{F}}$.
Nevertheless, the exact mean-field reduction of the dynamics at ${d \to \infty}$ allows us to do so without any loss of generality (see the discussion surrounding Eqs.~\eqref{eq-distr-cij-1}-\eqref{eq-distr-cij-2} on that point).

Finally, we recall the dynamical definition of the stress under a global shear strain (see Eq.~(49) in Ref.~\cite{agoritsas_maimbourg_zamponi_2019_JPhysA52_334001}):
\begin{equation}
 \hat{\sigma}^{\text{shear}}(t)
 \equiv \frac{\beta \sigma^{\text{shear}}(t)}{d \, \rho}
 = \frac{\widehat{\varphi}}2  \int^{\infty}_{-\infty}\!\!  \de h_0 \, \int \mathcal{D}g_1 \mathcal{D}g_2
 	 \, e^{h_0} g_{\text{in}}(h_0,g_1,g_2) \, g_1 g_2 \, \moy{\redv'(h(t))}_{h \vert h_0, g_1, g_2, \gamma(t)}
 \, .
\label{eq-DMFE-def-stress-global-shear}
\end{equation}
It consists in a statistical average of the force amplitude ${\moy{\redv'(h(t))}}$
under an accumulated strain ${\gamma(t)}$.
It is computed by averaging over the initial condition, \textit{i.e.}~the inital gap $h_0$ and the rescaled components of ${\hat{\rr}_0}$ in the shear plane ($g_{\mu}=\sqrt{d} \hat{r}_{0,\mu}$ for ${\mu=1,2}$).
This definition can be generalised to our new protocol,
%under constant random local displacements,
using first the translation stated in Eq.~\eqref{eq-translation-global-local-shear}
and secondly the decomposition of the gap ${h(t)}$ of Eq.~\eqref{eq-def-gap-in-AQRD}.
It gives the following expression:
\begin{equation}
\begin{split}
 \hat{\sigma}(t)
 \equiv \frac{\beta \sigma(t)}{d \, \rho}
 =& \frac{\widehat{\varphi}}2  \int^{\infty}_{-\infty}\!\!  \de h_0 \, e^{h_0} \, g_{\text{in}}(h_0) 
 	\int \mathcal{D}g_c \, \int_{\mathbb{R}} \de \tilde{c} \, \frac{e^{-\tilde{c}^2/(2\mathfrak{F})}}{\sqrt{2 \pi \mathfrak{F}}}
 	 \, \frac{\tilde{c}}{r_0} \,  g_c \, \moy{\redv'(h(t))}_{h \vert h_0, g_c, \tilde{c}/\rr_0,\gamma(t)}
 \\
 =& \frac{\widehat{\varphi}}2  \int^{\infty}_{-\infty}\!\!  \de h_0 \, e^{h_0} \, g_{\text{in}}(h_0) 
 	\int \mathcal{D}g_c \, \int \mathcal{D} \breve{c} \,
 	 \, \breve{c} \,  g_c \, \moy{\redv'(h(t))}_{h \vert h_0, g_c, \breve{c},\gamma_{\text{eff}}(t)}
 \, .
\end{split}
\label{eq-DMFE-def-stress-random-local-displ}
\end{equation}
The last equality emphasises once again the equivalence with global shear, upon the rescaling of the accumulated strain:
we can simply replace $g_1$ by ${\breve{c}=\tilde{c}/\ell}$, $g_2$ by $g_c$, and ${g_{\text{in}}(h_0,g_1,g_2)}$ by ${g_{\text{in}}(h_0)}$.
The latter holds by statistical isotropy.
In addition, it provides a quite intuitive definition of the stress in the system: it is  the statistical average over the scalar product, for a given interacting pair of particles, of the interparticle force at time $t$ and its assigned relative local displacement.
We have explicitly that:
\begin{equation}
 \frac{\tilde{c}}{r_0} \, g_c \, \moy{\redv(h(t))}
 = \frac{1}{r_0} \underbrace{\sqrt{d} \tilde{c} \, \hat{\cc}}_{\equiv \cc} \cdot \hat{\rr}_0 \, \moy{ v'(r(t))} \, \frac{\ell}{d}
 \approx \frac{1}{d} \moy{\cc_{ij} \cdot \nabla v \argp{\valabs{\rr_{ij} (t)}}}
 \, ,
 %= \frac{1}{d }\moy{\sum_{\mu=1}^{d} c_\mu \, \nabla_{\mu} v \argp{\valabs{\rr (t)}}}
\end{equation}
where we have reinstated the pair indices just to emphasise that it is computed over pairs.
The brackets denote here the dynamical average at fixed initial conditions.
This definition is in agreement with Eq.~(4) in Ref.~\cite{morse_roy_agoritsas_2020_Arxiv-2009.07706}, where the stress is computed as the scalar product in the ${Nd}$-dimensional phase space of the local relative displacement vector ${\vert \lbrace c_{ij} \rbrace \rangle}$ with the force vector ${\vert \lbrace F_{ij}(t) \rbrace \rangle}$ prescribed by the configuration at time $t$.
It enforces \emph{the notion that the stress is defined with respect to a given driving direction in configuration space}, as further discussed in Ref.~\cite{morse_roy_agoritsas_2020_Arxiv-2009.07706}\footnote{
More precisely, Eq.~(4) in Ref.~\cite{morse_roy_agoritsas_2020_Arxiv-2009.07706} defines the stress as ${\sigma \propto \sum_{i=1}^N {\bf F}_i \cdot \cc_i}$, with a normalisation with respect to system size and number of particles $N$ that we skip for now. This definition relies on the forces acting on each particle ${\lbrace {\bf F}_i \rbrace}$ and the associated local displacement ${\lbrace \cc_i \rbrace}$. Our definition relies instead on the forces between pairs of particles ${\lbrace {\bf F}_{ij} \rbrace}$ and the associated relative displacements ${\lbrace \cc_{ij} \rbrace}$. These two definitions are in fact strictly equivalent, at any given time $t$:
$$
\frac12 \sum_{i,j =1}^N {\bf F}_{ij} \cdot \cc_{ij}
=\sum_{i,j =1}^N {\bf F}_{ij} \cdot \argp{\cc_i - \cc_j}
\stackrel{[{\bf F}_{ij} = -{\bf F}_{ji}]}{=}
	\frac12 \sum_i \underbrace{\sum_{j (\neq i)} {\bf F}_{ij}}_{= {\bf F}_i} \cdot \cc_i
	+ \frac12 \sum_j \underbrace{\sum_{i (\neq j)} {\bf F}_{ji}}_{= {\bf F}_j} \cdot \cc_j
= \sum_i {\bf F}_{i} \cdot \cc_i \, .
$$
}.

%_____________________________________________________________________________________________________
\section{Quasistatic driving of glassy states}
\label{section-quasistatics-glassy-states}

We now focus on the quasistatic driving of glassy states,
meaning that we start from an equilibrium initial condition below the dynamical transition temperature~\cite{book_parisi_urbani_zamponi_2020}.
We aim in particular at comparing the AQRD and AQS protocols, as discussed in Ref.~\cite{morse_roy_agoritsas_2020_Arxiv-2009.07706}.
%
%and we restrict ourselves to the case an replica-symmetric (RS) initial equilibrium.
%
Such quasistatic driving can be investigated directly with static approaches~\cite{rainone_2015_PhysRevLett114_015701,rainone_urbani_2016_JStatMech2016_053302,biroli_urbani_2016_NatPhys12_1130,
urbani_zamponi_2017_PhysRevLett118_038001,biroli_urbani_2018_SciPostPhys4_020,altieri_2019_PhysRevE100_032140},
and in Refs.~\cite{agoritsas_maimbourg_zamponi_2019_JPhysA52_144002,agoritsas_maimbourg_zamponi_2019_JPhysA52_334001} we provided a dynamical derivation of such `state-following protocols'.
Because of the equivalence with the case of global shear, that we have established in the previous section \emph{at the level the dynamics}, we can straightforwardly adapt the results presented in Ref.~\cite{agoritsas_maimbourg_zamponi_2019_JPhysA52_334001} (specifically in Sec.~V),
using the notations and definitions of the review book~\cite{book_parisi_urbani_zamponi_2020}.
However, we emphasise that our formalism allows us to include more generally the possibility of a finite temperature, directly in the many-body dynamics given by Eqs.~\eqref{eqC3:GENLang-shear-dynamics-AQRD}-\eqref{eqC3:GENLang-shear-noise}.
We are thus not restricted to the pure athermal case, even though we refer by convention to the AQRD and AQS protocols.

In the `solid' glassy phase we have at short timescales equilibrium-like fluctuations (\textit{i.e.}~satisfying fluctuation-dissipation relations), within the local metabasins of a disordered energy landscape.
The statistics of these metabasins depend on the preparation protocol of the system (we assume it to be replica-symmetric, for now).
By definition, quasistatic driving amounts to slowly modifying this landscape while always enforcing equilibrium at short timescales.
In practice, we substitute such a timescale separation ansatz in our dynamical mean-field equations (DMFE) and we assume that the mean-square-displacements (MSD) functions saturate to a finite plateau at long times, which can in turn be interpreted as the typical size of the local metabasins. %(note that we assume a replica-symmetric landscape).
The set of self-consistent equations for this plateau value has a solution as long as the system behaves as a `solid': this solution eventually breaks down when either increasing the temperature, lowering the packing fraction, shearing too much, or increasing the variance of random forces.
Note that we focus here on the RS phase, but the system might undergo a transition to a full replica-symmetric phase \cite{rainone_2015_PhysRevLett114_015701,rainone_urbani_2016_JStatMech2016_053302} before yielding `for good'.

%% VECTORIAL
%
We consider the case of a `strain' ${\gamma}$ smoothly applied over a finite timescale $\tau$, such that at long times the affine motion simply becomes ${\rr_0'(t) \to \rr_0 + \gamma \cc \equiv \hat{S}_{\gamma {\bf c}} \rr_0}$.
We assume the same ansatz as in Eq.~(51)~\cite{agoritsas_maimbourg_zamponi_2019_JPhysA52_334001},
which thus leads to the same restricted equilibrium distributions as in Eq.~(55)~\cite{agoritsas_maimbourg_zamponi_2019_JPhysA52_334001},
replacing only ${\hat{S}_{\gamma}}$ in AQS by ${\hat{S}_{\gamma {\bf c}}}$ in AQRD.
In addition to integrating over the initial condition ${\int \de \rr_0 \, g_{\text{in}}(\rr_0)}$, we have to average over the local random displacements ${\int \de {\bf c} \, \bar{\mathcal{P}}({\bf c}) \, (\dots)}$.
We complement the ansatz by assuming that we have at long times the MSD plateaus $D_r$ and $D$, with the following definitions (recalled from Eq.~(56)~\cite{agoritsas_maimbourg_zamponi_2019_JPhysA52_334001}):
%and complemented by Eq.~(143)~\cite{agoritsas_maimbourg_zamponi_2019_JPhysA52_144002} for the addition of random forces ${f_0^2>0}$):
\begin{equation}
\label{eq:MSDslongtime}
 D_r 
 	= \frac1d \lim_{t\to\infty} \moy{ \valabs{ \xx(t) - \xx(0) }^2} 
 	%= \frac1d \overline{ \moy{ \valabs{ \uu }^2 }_{\rm req}}
 	%= \frac{T}{k_{\rm eff}} + \frac{M_\infty \meevid{+ f_0^2}}{k_{\rm eff}^2} 
 \, , \quad
 D
	= \frac1d \lim_{t\to\infty} \lim_{|t-s|\to\infty}  \moy{ \valabs{ \xx(t) - \xx(s) }^2 } 
	%= \frac2d \overline{ \argc{ \moy{ \valabs{ \uu }^2}_{\rm req} - \valabs{ \moy{ \uu }_{\rm req} }^2 } }
	%=  \frac{2T}{k_{\rm eff}}
 \, , \quad
 A \equiv 2 D_r - D
 \, .
\end{equation}
%
%Physically, we recall that
%$D_r$ is the MSD of each particle with respect to its initial position,
%$D$ the MSD of the inter-particle distance fluctuation,
%and $M_\infty$ the long-time force-force correlator (and characterises the metabasins distribution). 
%${A= \overline{\valabs{ \moy{ \uu }_{\rm req} }^2} = 2 D_r - D = 2 (M_\infty \meevid{+f_0^2})/k_{\rm eff}^2}$.
%
We assume that we start from an RS equilibrium at inverse temperature ${\beta_0}$, that we are in contact with a thermal bath at inverse temperature ${\beta}$, and for completeness we allow for having constant random forces with ${\Gamma_C(t,t')=f_0^2}$ in Eq.~\eqref{eqC3:GENLang-shear-noise}.
These quantities have to satisfy the following set of equations, given first in their high-dimensional vectorial form:
\begin{equation}
\label{eq-state-following-equa-AQRD}
\begin{split}
 & \frac{1}{D} - \frac{A}{D^2}
 	%&= \frac{1}{2T} \argc{ k_{\rm eff} - \frac{1}{T} \argp{M_\infty \meevid{+ f_0^2}}}
	%\\
	= - \frac{\beta^2 f_0^2}{2}-\frac{\rho}{d} \int \de \rr_0 \, e^{-\beta_0 v(\rr_0)} \int \de \cc \, \bar{\mathcal{P}}(\cc) \, e^{\frac{A}{2} \nabla^2 }
	\argc{ \frac{ \frac{\nabla^2}2 e^{\frac{D}{2} \nabla^2}  e^{-\beta v(\hat{S}_{\gamma {\cc}}  \rr_0 )} }
{ e^{\frac{D}{2}\nabla^2} e^{-\beta v(\hat{S}_{\gamma {\cc}}  \rr_0 )}}  }
 \, ,
 \\
 & \frac1{D}
	%=	\frac{k_{\rm eff}}{2T}
	= -\frac{\rho}{d} \int \de \rr_0 \, e^{-\beta_0 v(\rr_0)} \int \de {\cc} \, \bar{\mathcal{P}}({\cc}) \, e^{\frac{A}{2} \nabla^2   } \frac{\nabla^2}2\log
			\argc{ e^{\frac{D}2 \nabla^2} e^{-\beta v(\hat{S}_{\gamma {\cc}}  \rr_0)} }
 \, .
\end{split}
\end{equation}
where we used the compact notation
${e^{\frac{\tilde{\Delta}}2 \nabla^2} f(\rr)= \int \de \xx \, \frac{e^{- \frac{\xx^2}{2 \tilde{\Delta}}}}{(2\pi \tilde{\Delta})^{d/2}} f(\rr + \xx)}$.
The corresponding glassy free energy ${{\rm f}_g}$, which is the quantity that would be computed and studied  using replic{\ae} in statics approaches~\cite{book_parisi_urbani_zamponi_2020}, is
\begin{equation}
\label{eq-state-following-free-energybis-AQRD}
 - \frac{2}{d} \beta {\rm f}_g
 	= \argc{ 1 + \log(\pi D) + \frac{A}{D} }
 		+ \frac{\beta^2 f_0^2 D}{2}
 		+ \frac{\rho}{d} \int \de \rr_0 \, e^{\beta_0 v(\rr_0)}  \int \de \cc \, \bar{\mathcal{P}}(\cc)  \, e^{\frac{A}2 \nabla^2} \log \argc{ e^{\frac{D}2 \nabla^2} e^{-\beta v(\hat{S}_{\gamma \cc} )} }
\, .
\end{equation}
One can check that we can recover in particular Eqs.~\eqref{eq-state-following-equa-AQRD} from the extremalization conditions ${\partial_\Delta {\rm f}_g =0}$ and ${\partial_A {\rm f}_g =0}$, as it should be.

%% SCALAR
%
For explicit computations of observables, such as the stress-strain curves, we need to work with the scalar formulation of the free energy, which is much more user-friendly and essentially involves scalar Gaussian convolutions.
It can be obtained either starting from Eq.~\eqref{eq-state-following-free-energybis-AQRD}, or from the scalar mean-field dynamics with the scalar counterpart of the quasistatic ansatz.
Thereafter we directly use the expressions from the review book~\cite{book_parisi_urbani_zamponi_2020}, starting from the definitions of the rescaled MSD functions
${\Delta_r = \frac{d^2}{\ell^2} D_r}$, ${\Delta = \frac{d^2}{\ell^2} D}$, and ${\mathcal{A} = 2 \Delta_r - \Delta}$.
We also consider the rescaled random forces variance ${\hat{f}_0^2 = \frac{\ell^2}{2 d^2} f_0^2}$, using the same rescaling as for ${\mathcal{G}_C(t,s)=\frac{\ell^2}{2 d^2} \Gamma_C(t,s)}$ (given after Eq.~\eqref{eq-def-longitudinal-motion-stoch-process-AQRD}).
The glass free energy takes the now standard form~\cite{book_parisi_urbani_zamponi_2020}:
\begin{equation}
  - \frac{2}{d} \beta {\rm f}_g
 	= \argc{1 + \log \argp{\pi \ell^2 \Delta /d^2} + \frac{2 \Delta_r - \Delta}{\Delta}}
 		+ \beta^2 \hat{f}_0^2 \Delta
 		+ \widehat{\varphi} \int_{\mathbb{R}} \de h_0 \, e^{h_0} \, \underbrace{q_{\gamma}\argp{2 \Delta_r - \Delta, \beta_0 ; h_0}}_{\text{only dependence on $\gamma$}} \, \log q \argp{\Delta, \beta;h_0}
 \, ,
\label{eq-fg-scalar-shear}
\end{equation}
with $\widehat{\varphi}$ the rescaled packing fraction as defined after Eq.~\eqref{eq-def-kernels-high-dim-gap-AQRD}, and the following definitions:
\begin{eqnarray}
 q \argp{\Delta, \beta; h_0}
 	&\equiv& \int_{\mathbb{R}} \de h \, e^{-\beta \redv(h)} \frac{e^{-\frac{(h-h_0-\Delta/2)^2}{2 \Delta}}}{\sqrt{2 \pi \Delta}}
 \, ,
\label{eq-fg-scalar-shear-q-function-generic}
 \\
 q_{\gamma}^{\text{AQS}} \argp{\widetilde{\Delta}, \beta_0 ; h_0}
 	&\equiv& \int \de g_2 \, \frac{e^{-g^2/2}}{\sqrt{2 \pi}} \, q \argp{\widetilde{\Delta} + \gamma^2 g_2^2, \beta_0 ; h_0}
 	\equiv \int \mathcal{D} g_2 \, q \argp{\widetilde{\Delta} + \gamma^2 g_2^2, \beta_0 ; h_0}
 	\, ,
\label{eq-fg-scalar-shear-q-function-AQS}
 \\
 q_{\gamma}^{\text{AQRD}}\argp{\widetilde{\Delta}, \beta_0 ; h_0}
 	&\equiv& \int \de \tilde{c} \, \frac{e^{-\tilde{c}^2/(2 \mathfrak{F})}}{\sqrt{2 \pi \mathfrak{F}}} \, q \argp{\widetilde{\Delta} + \gamma^2 \frac{\tilde{c}^2}{\ell^2}, \beta_0 ; h_0}
	= \int \de \breve{c} \, \frac{e^{-\breve{c}^2/2}}{\sqrt{2 \pi}} \, q \argp{\widetilde{\Delta} + \underbrace{\gamma^2 \frac{\mathfrak{F}}{\ell^2}}_{\gamma_\text{eff}^2} \breve{c}^2, \beta_0 ; h_0}
 \, .
\label{eq-qgammaAQRD}
\end{eqnarray}
In that case, we see that the free energy (from which all static observables can be derived) is strictly equivalent for AQS and AQRD \emph{provided that we rescale accordingly the accumulated strain}, specifically by replacing $\gamma$ by ${\gamma_{\text{eff}}= \gamma \sqrt{\mathfrak{F}}/\ell}$.
This is a direct consequence of the equivalence between global shear and constant random local displacements
--~that we have shown in Sec.~\ref{section-recalling-DMFE} to be valid for the whole dynamics~--
which had to hold in particular for quasistatic drivings.

All those quantities can be computed at least numerically, for a given interaction potential ${v(r) = \redv(h)}$.
In practice, the set of equations for $\arga{\Delta,\mathcal{A}=2\Delta_r - \Delta}$ is obtained by extremalization of the free energy~\eqref{eq-fg-scalar-shear}, with ${\partial_{\Delta} {\rm f}_g =0}$ and ${\partial_{\Delta_r} {\rm f}_g =0}$, yielding the following equations:
\begin{equation}
\begin{split}
 & \frac{1}{\Delta} - \frac{\mathcal{A}}{\Delta^2}
 	= -\beta^2 \hat{f}_0^2 - \widehat{\varphi} \int_{\mathbb{R}}\de h_0 \, e^{h_0} \, q_{\gamma} \argp{\mathcal{A}, \beta_0 ; h_0} \frac{\partial_\Delta q \argp{\Delta, \beta ; h_0}}{q \argp{\Delta, \beta ; h_0}}
 	%= -\frac{\beta^2 \hat{f}_0^2}{2} - \widehat{\varphi} \int_{\mathbb{R}}\de h_0 \, e^{h_0} \, q_{\gamma} \argp{2 \Delta_r - \Delta, \beta_0 ; h_0} \frac{\partial_\Delta q \argp{\Delta, \beta_0 ; h_0}}{q \argp{\Delta, \beta_0 ; h_0}}
 \, ,
 \\
 & \frac{1}{\Delta}
 	= - \widehat{\varphi} \int_{\mathbb{R}}\de h_0 \, e^{h_0} \, \partial_{\mathcal{A}} q_{\gamma} \argp{\mathcal{A}, \beta_0 ; h_0} \, \log q \argp{\Delta, \beta ; h_0}
 	%= - \frac{\widehat{\varphi}}{2} \int_{\mathbb{R}}\de h_0 \, e^{h_0} \, \partial_{\Delta_r} q_{\gamma} \argp{2 \Delta_r - \Delta, \beta_0 ; h_0} \, \log q \argp{\Delta, \beta_0 ; h_0}
 \, .
\end{split}
\label{eq-MSD-scalar-complete}
\end{equation}
If there exists a solution to these equations, this means that we are in the solid phase (or at least that such a solid phase can be sustainable).
Once we have self-consistently determined the values of ${\arga{\Delta,\mathcal{A}}}$ or ${\arga{\Delta,\Delta_r}}$ satisfying these equations,  we can compute observables such as the shear stress or the pressure~\cite{book_parisi_urbani_zamponi_2020}.
The principal interest of including the case with constant random forces is that Eq.~\eqref{eq-MSD-scalar-complete} can then be used more generally to study not only `strain'-controlled protocols
(controlling ${\gamma(t)}$ at ${\hat{f}_0^2=0}$),
but also `stress'-controlled counterparts
(controlling ${f_0^2}$ at ${\gamma(t)=0}$).
However, this issue will be further discussed in future work.
In the next section, we focus on the former case, which was the case under study in Ref.~\cite{morse_roy_agoritsas_2020_Arxiv-2009.07706}, for which the corresponding stress is computed as ${\sigma = \partial_{\gamma} {\rm f}_g}$ (see Eq.~(10.15) of~\cite{book_parisi_urbani_zamponi_2020}).

%_____________________________________________________________________________________________________
\section{AQRD \textit{versus} AQS stress-strain curves and elastic modulus}
\label{section-AQRD-stress-strain-elastic-modulus}

We have established in Sec.~\ref{section-recalling-DMFE} the equivalence between global shear and constant random local displacements at the level of the dynamics, which includes the quasistatic case as explicitly shown in Sec.~\ref{section-quasistatics-glassy-states}.
The implications on the quasistatic stress-strain curves and their associated elastic moduli, in the comparison between the AQRD or AQS protocols, are quite straightforward to obtain, as we discuss below.
%from the glass free energy given by Eqs.~\eqref{eq-fg-scalar-shear}-\eqref{eq-qgammaAQRD}.
%
We recall that our formalism allows for a finite temperature, so our predictions are not restricted to the athermal case.
Note however that we use thereafter the AQS/AQRD abbreviation by convention and in direct reference to the results announced in Ref.~\cite{morse_roy_agoritsas_2020_Arxiv-2009.07706}.

For a quasistatic global shear in the pre-yielding regime, starting from a RS equilibrium glass, the stress can be computed by taking the derivative of the glass free energy given by Eqs.~\eqref{eq-fg-scalar-shear}-\eqref{eq-qgammaAQRD} with respect to the `true' strain, ${\sigma_{\text{AQS}}(\gamma)=\partial_{\gamma} {\rm f}_g^{\text{AQS}}(\gamma)}$~\cite{book_parisi_urbani_zamponi_2020}.
This definition has been used for instance to compute the quasistatic mean-field stress-strain curves given in Refs.~\cite{rainone_2015_PhysRevLett114_015701,rainone_urbani_2016_JStatMech2016_053302,
urbani_zamponi_2017_PhysRevLett118_038001,altieri_2019_PhysRevE100_032140},
for sheared hard-sphere systems.
Consequently, their AQRD counterparts are simply obtained by the following rescaling:
%Regarding the stress-strain curves, the stress can be obtained by taking the derivative of the glass free energy with respect to the `true' strain, ${\sigma_{\text{shear}}(\gamma)=\partial_{\gamma} {\rm f}_g^{\text{shear}}(\gamma)}$, which gives rather straightforwardly for the AQRD case:
\begin{equation}
\gamma_\text{eff} = \gamma \sqrt{\mathfrak{F}}/\ell
\quad \Rightarrow \quad
 \sigma_{\text{AQRD}}(\gamma)
 	\equiv \partial_\gamma {\rm f}_g^{\text{AQRD}} \argp{\gamma}
 	= \partial_\gamma {\rm f}_g^{\text{AQS}} \argp{\gamma_{\text{eff}}}
 	= \frac{\sqrt{\mathfrak{F}}}{\ell} \, \frac{\partial {\rm f}_g^{\text{AQS}} \argp{\gamma_{\text{eff}}}}{\partial {\gamma_{\text{eff}}} } 
 	\equiv \frac{\sqrt{\mathfrak{F}}}{\ell} \, \sigma_{\text{AQS}}(\gamma_{\text{eff}})
 	%= \frac{\sqrt{\mathfrak{F}}}{\ell} \, \sigma_{\text{AQS}}(\gamma \sqrt{\mathfrak{F}}/\ell)
 \, .
\label{stress-strain-curve-AQRD}
\end{equation}
As for the elastic modulus at zero strain, using the definition given by Eq.~(10.18) in~\cite{book_parisi_urbani_zamponi_2020}, it is simply given by:
\begin{equation}
 \mu_{\text{AQS}}
 	\equiv \frac{\de \sigma_{\text{AQS}}(\gamma)}{\de \gamma}\Big\vert_{\gamma=0}
 \quad \Rightarrow \quad
 \mu_{\text{AQRD}}
 	\equiv \frac{\de \sigma_{\text{AQRD}}(\gamma)}{\de \gamma}\Big\vert_{\gamma=0}
 	= \frac{\mathfrak{F}}{\ell^2} \, \frac{\de \sigma_{\text{AQS}}(\gamma_{\text{eff}})}{\de \gamma_{\text{eff}}}\Big\vert_{\gamma_{\text{eff}}=0}
 	= \frac{\mathfrak{F}}{\ell^2} \mu_{\text{AQS}}
 \, ,
\label{elastic-modulus-zero-strain-AQRD}
\end{equation}
where ${\mu_{\text{AQS}}}$ is fixed by
\textit{(i)}~the rescaling packing fraction ${\widehat{\varphi}}$,
\textit{(ii)}~the initial and final inverse temperatures ${\arga{\beta_0,\beta}}$,
and of course \textit{(iv)}~the specific interaction potential ${v(r)=\redv(h)}$ with its typical interaction length ${\ell}$ fixing the units of length.
Eq.~\eqref{elastic-modulus-zero-strain-AQRD} is particularly interesting since it predicts the ratio $\kappa$ of the initial elastic moduli to be directly given by the unitless variance ${\mathfrak{F}/\ell^2}$.
These predictions can be rewritten as follows,
denoting by $\tilde{\gamma}$ the `random strain' controlled in AQRD
and $\gamma_{\text{shear}}$ the corresponding AQS strain,
in order to match the notations used in Ref.~\cite{morse_roy_agoritsas_2020_Arxiv-2009.07706} (see its Eq.~(7)):
\begin{equation}
 \kappa \equiv \frac{\mu_{\text{AQRD}}}{\mu_{\text{AQS}}} = \frac{\mathfrak{F}}{\ell^2}
 \, , \quad
 \gamma_{\text{shear}} = \tilde{\gamma} \sqrt{\kappa}
 \, , \quad
 \sigma_{\text{AQS}} = \sigma_{\text{AQRD}}(\tilde{\gamma}) / \sqrt{\kappa}
 \, .
\label{eq-prescription-rescaling-stress-strain-curves-AQRD-AQS}
\end{equation}
These relations prescribe how one should rescale the AQRD mean-field stress-strain data to make them collapse on their AQS counterpart, and the other way around how to generate AQRD curves from an AQS one.
These predictions, obtained in the infinite-dimensional limit, might extend to lower dimensions as far as mean-field-like quantities are considered.
They have been tested in Ref.~\cite{morse_roy_agoritsas_2020_Arxiv-2009.07706} on two-dimensional numerical simulations, and the quantitative agreement was remarkably good for the stress-strain curves, the distributions of elastic moduli, of strain-step between stress drops (\textit{i.e.}~of the distance between two successible saddle points of the potential energy landscape), and of stress drops themselves.

\begin{figure}[h]
\begin{center}
\includegraphics[width=0.9\textwidth]{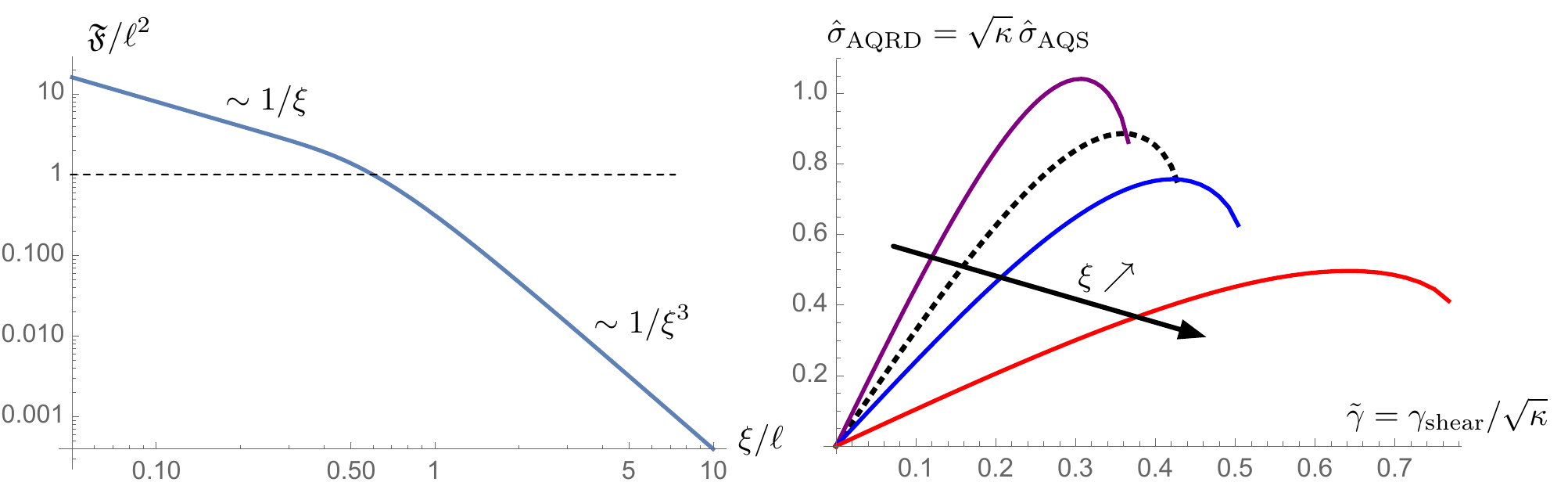}
\caption{
\textit{Left:}~Variance of relative displacements ${\mathfrak{F}}$ as a function of the correlation length $\xi$ (both rescaled by the typical interparticle distance $\ell$),
assuming a Gaussian correlation as in Eq.~\eqref{eq-mathfrak-F-explicit-Gaussian_fxi} with an overall amplitude ${\Xi=1}$. This amplitude can be tuned to shift the crossover in the scaling of $\xi$, with respect to the value ${\mathfrak{F}/\ell^2=1}$ which indicates the collapse onto the AQS case.
\textit{Right:}~Rescaled AQRD stress $\hat{\sigma}_{\text{AQRD}}$ (as defined in Eq.~\eqref{eq-DMFE-def-stress-random-local-displ}) as a function of the `random' strain ${\tilde{\gamma}}$, generated by rescaling of the AQS stress-strain curve in dotted black (reproduced from Fig.~2 in Ref.~\cite{rainone_2015_PhysRevLett114_015701}, for hard spheres at a packing fraction ${\widehat{\varphi}=6}$) according to Eq.~\eqref{eq-prescription-rescaling-stress-strain-curves-AQRD-AQS}.
The three continuous curves correspond to ${\xi/\ell \in \lbrace 0.5, 0.7, 1 \rbrace}$ (respectively purple, blue and red), as indicated by the black arrow for increasing $\xi$.
They correspond to ${\mathfrak{F}/\ell^2 \in \lbrace 1.38 , 0.73 , 0.31 \rbrace}$ for the amplitude fixed at ${\Xi=1}$.
Note that the overall amplitude $\Xi$ is also a control parameter of our protocol, whereas in Ref.~\cite{morse_roy_agoritsas_2020_Arxiv-2009.07706} its value is fixed by the imposed normalisation ${\langle c \vert c \rangle=1}$.
% From Mathematica {1., 1.37981, 0.728984, 0.313943}
}
\label{fig-rescaled-stress-strain-curves}
\end{center}
\end{figure}

According to these predictions, ${\mu_{\text{AQRD}} \propto \mathfrak{F} \, \mu_{\text{shear}}}$.
If we assume in particular ${f_{\xi}(x)}$ to be a normalised Gaussian function,
as in Eq.~\eqref{eq-mathfrak-F-explicit-Gaussian_fxi},
we then expect the elastic modulus to have a crossover depending on the ratio ${\ell/\xi}$,
with ${\mathfrak{F} \sim 1/\xi}$ at ${\ell/\xi \gg 1}$
and ${\mathfrak{F} \sim 1/\xi^{3}}$ at ${\ell/\xi \ll 1}$.
The latter case corresponds to a global shear, where $\xi$ is of the order of the system size~\cite{morse_roy_agoritsas_2020_Arxiv-2009.07706}.
In both cases, this implies that the elastic modulus decreases with increasing $\xi$, as we numerically observe and physically expect: it is more efficient to deform a glass with local displacements which are more disordered, \textit{i.e.}~with a smaller correlation length.
Less obvious is that a larger correlation length between individual displacements implies a smaller variance of the relative displacements, \textit{i.e.}~$\mathfrak{F}$ must decrease with increasing $\xi$.
This crossover is illustrated in Fig.~\ref{fig-rescaled-stress-strain-curves}:
on the left for
${\mathfrak{F}} \propto \kappa \equiv \mu_{\text{AQRD}}/\mu_{\text{AQS}}$,
and on the right for a set of stress-strain curves,
obtained by rescaling ${\sigma_{\text{AQS}}(\gamma)}$ computed for hard spheres in Ref.~\cite{rainone_2015_PhysRevLett114_015701} (see its Fig.~2, for a packing fraction ${\widehat{\varphi}=6}$).
Because of the power-law dependence of $\mathfrak{F}$ on $\xi$, these curves can easily be shifted by several order of magnitude (including their ending point, which is related to the yielding point).

%_____________________________________________________________________________________________________
\section{Conclusion}
\label{section-discussion-conclusion}

We have established the exact equivalence between global shear and a local forcing
--~in the form of imposed constant random local displacements~--
for infinite-dimensional particle systems with pairwise interactions.
%with a finite spatial correlation length.
%
By adapting the derivation for global shear detailed in Ref.~\cite{agoritsas_maimbourg_zamponi_2019_JPhysA52_334001},
we have shown that this equivalence holds at the level of the full mean-field dynamics upon a simple rescaling of the accumulated effective strain (see Sec.~\ref{section-recalling-DMFE}).
This statement holds in particular for quasistatic drivings (in Sec.~\ref{section-quasistatics-glassy-states}), and culminates in the prediction that the AQS and AQRD stress-strain curves and their initial elastic modulus can be collapsed on each other via Eq.~\eqref{eq-prescription-rescaling-stress-strain-curves-AQRD-AQS} (see Sec.~\ref{section-AQRD-stress-strain-elastic-modulus}).

%-----------

The key parameter of this equivalence is the unitless variance ${\mathfrak{F}/\ell^2}$ of the relative local displacements
--~restricted on interacting pairs of particles~--
which keeps an explicit dependence on the spatial correlation of the local forcing.
An increasing correlation length implies a decreasing variance: a coordinated forcing deforms an amorphous material less efficiently than a completely uncorrelated random forcing, in the sense that it corresponds to a smaller effective accumulated strain ${\gamma_{\text{eff}} = \gamma \sqrt{\mathfrak{F}}/\ell}$.
In our derivation, we first assume a Gaussian distribution for the local deformation field.
In addition, for explicit computations we consider a Gaussian two-point correlator ${f_\xi}$ (in Eq.~\eqref{eq-distr-cij-2})
%${f_\xi(\valabs{\xx - \xx'})}$,
for which we predict a crossover from ${\mathfrak{F}(\xi) \sim 1/\xi}$ at ${\xi/\ell \ll 1}$ to ${\sim 1/\xi^3}$ in the opposite limit.
The latter is relevant to compare to the global shear case, which by construction has a value for $\xi$ comparable to the system size (see Sec.~\ref{section-settings-global-vs-local-shear-strain}).
Remarkably, except for these scalings, our construction is not specific to these assumptions.
%thanks to the infinite-dimensional limit.
%
It relies indeed on the conjunction of two features which are exact in the infinite-dimensional limit:
\textit{(i)}~different pairs of particles effectively do not interact,
in the sense that their contribution becomes irrelevant in path-integral statistical averages (see for instance Ref.~\cite{agoritsas_maimbourg_zamponi_2019_JPhysA52_144002}),
so the effective mean-field dynamics is controlled by single-pair statistics; %of relative displacements;
\textit{(ii)}~the statistics of relative displacements tends to a Gaussian distribution, uncorrelated between different pairs and fully characterised by the variance on a given pair.
We could consequently consider alternative deformation fields, as for instance wave-like patterned field as in Ref.~\cite{morse_roy_agoritsas_2020_Arxiv-2009.07706}.
Their sole relevant feature for the mean-field dynamics in infinite dimension will always be their variance $\mathfrak{F}$, albeit with a different functional dependence on $\xi$.

%-----------

A pending issue regarding such infinite-dimensional results is their relevance regarding systems in low dimensions.
As emphasised in the introduction, this limit has the advantage that
exact mean-field predictions might be within reach,
%one might be able to obtain exact mean-field predictions,
while on the other hand important physics due to spatial correlations might be completely washed out.
Since this work has been done in parallel to the numerical study presented in Ref.~\cite{morse_roy_agoritsas_2020_Arxiv-2009.07706},
we were able to directly address this issue by comparing our mean-field predictions~\eqref{eq-prescription-rescaling-stress-strain-curves-AQRD-AQS} to 2D numerics under either AQRD or AQS protocols.
Beyond qualitatively similar behaviours, we found a remarkably good quantitative agreement regarding the stress-strain curves and the avalanches statistics.
First, we could collapse the different stress-strain curves and the distribution of stress drops and strain steps via a rescaling controlled by the ratio of initial elastic moduli ${\kappa=\mu_{\text{AQRD}}(\xi)/\mu_{\text{AQS}}}$.
Secondly we could measure the unitless variance ${\mathfrak{F}(\xi)/\ell^2}$ and show that it coincides with $\kappa$ (as predicted in infinite dimension) for wave-like patterned displacement field, and also for the Gaussian random displacements that we considered here.
For the latter, quantitative discrepancies nevertheless appear when the correlation length $\xi$ decreases, hinting at the increasing role of spatial correlations in these low-dimensional systems.

%-----------

Although both Ref.~\cite{morse_roy_agoritsas_2020_Arxiv-2009.07706} and our Sec.~\ref{section-quasistatics-glassy-states}-\ref{section-AQRD-stress-strain-elastic-modulus} focus on quasistatic driving, our statement about the exact equivalence between global shear and local forcing holds for the whole dynamics
of infinite-dimensional particle systems with pairwise interactions.
It holds in particular for a finite, possibly time-dependent, shear rate ${\dot{\gamma}(t)}$.
We keep the investigation of the resulting rheological behaviours for future work.
For now we only emphasise that in that case, the local forcing on each particle is strictly speaking a local \emph{force}, equal to ${\zeta \dot{\gamma}(t) {\bf c}_i}$ in Eq.~\eqref{eqC3:GENLang-shear-dynamics-AQRD}.
It thus corresponds to having a local force field correlated in space and with an amplitude proportional to the shear rate.
In low dimensions, the similarity between the rheology at constant shear rate of 2D particle systems, either under global shear or with self-propulsion, has been addressed in Ref.~\cite{liao_xu_2018_SoftMatter14_853}, but assuming a fixed norm of the local forces:
a missing ingredient to quantitatively connect these two types of rheology might be to allow for a spatial correlation of the local forcing.
Remarkably, the idea of applying such a local forcing to mimic the rheology at a constant shear rate has already been considered in Ref.~\cite{berthier_2000_PhysRevE61_5464}.
This was done by adding a non-conservative force on the spherical $p$-spin model, and then invoking a so-called `schematic' modelling of mode-coupling relations.
In that respect, our results allow us to bypass the possible objections regarding either the analogy between the driven $p$-spin model and the rheology of dense interacting-particle systems, or the limitations of the mode-coupling approximation: again, in our infinite-dimensional framework, dynamics under global shear or random local forcing are strictly equivalent.

%-----------

From a technical point of view, we adapted the previous derivation presented in Ref.~\cite{agoritsas_maimbourg_zamponi_2019_JPhysA52_334001}, where we had used a so-called `dynamical cavity' approach in order to obtain the dynamical mean-field equations under a global shear protocol.
This approach consists, in a nutshell, in performing a self-consistent Taylor expansion at leading order in the infinite-dimensional limit.
A complementary but technically more involved derivation was earlier presented in the companion paper~\cite{agoritsas_maimbourg_zamponi_2019_JPhysA52_144002} --focused on the isotropic case--
based on a super-symmetric path-integral formulation and its infinite-dimensional saddle point.
Here our new derivation focuses on the implementation of spatial correlations in the local forcing, either on local displacements (key for the  quasistatic limit of vanishing shear rate) or on local forces (at any finite shear rate ${\dot{\gamma}(t)}$ in Eq.~\eqref{eqC3:GENLang-shear-dynamics-AQRD}).
The concrete translation of such spatial correlations in a path-integral formulation would further support our dynamical cavity derivation, properly justifying the underlying assumptions as exact features of the ${d \to \infty}$ path-integral saddle-point.
Moreover, it would allow for direct connections to other models with spatially-correlated disorder, such as elastic random manifolds for which disorder correlator is the key quantity to follow upon functional renormalisation approaches~\cite{chauve_2000_ThesePC_PhysRevB62_6241}.

%-----------

These results support the physical picture that there is indeed a proper equivalence between global shear and local forcing, regarding the statistical sampling of the configurational phase space, at least as long as we focus on mean-field metrics.
A promising perspective to this work would be to challenge this picture in other driven disordered systems,
and to systematically disentangle such mean-field behaviour from spatial-correlation effects, depending on the specific nature of the driving.

%_____________________________________________________________
\section*{Acknowledgments}

I would like to warmly thank Peter Morse, and also Eric Corwin, Lisa Manning, Sudeshna Roy and Ethan Stanifer, for the discussions on our common work~\cite{morse_roy_agoritsas_2020_Arxiv-2009.07706} which motived this whole computation, as well as Francesco Zamponi and Ada Altieri for their helpful insight on the infinite-dimensional limit.
This work was supported by the Swiss National Science Foundation under the SNSF Ambizione Grant PZ00P2{\_}173962.

%_____________________________________________________________
%% Bibliography/References

%\bibliographystyle{unsrt}
\bibliographystyle{plain_url}
%\bibliography{BiblioEli}

%_____________________________________________________________

\end{document}